%% file: main.tex
\documentclass[10pt,twocolumn,letterpaper]{article}
\usepackage{iccv}
\toggletrue{iccvfinal}
\usepackage{subcaption}
\usepackage{graphicx}
\usepackage{multicol}
\usepackage{multirow}
\usepackage{float}
\usepackage{booktabs}
\usepackage{caption}
\usepackage{arydshln} 
\usepackage{wrapfig} 
\usepackage{placeins}
\usepackage{indentfirst}       
\usepackage[pagebackref,breaklinks,colorlinks,allcolors=iccvblue]{hyperref}
\usepackage{xr-hyper}
\usepackage[capitalize]{cleveref}
\usepackage{lipsum}

\newcommand\blfootnote[1]{%
  \begingroup
  \renewcommand\thefootnote{}\footnote{#1}%
  \addtocounter{footnote}{-1}%
  \endgroup
}


\definecolor{iccvblue}{rgb}{0.21,0.49,0.74}

\title{AJAHR: Amputated Joint Aware 3D Human Mesh Recovery}

\author{
Hyunjin Cho\textsuperscript{1,3,*} \quad
Giyun Choi\textsuperscript{1,*} \quad
Jongwon Choi\textsuperscript{1,2,$\dag$} \\[1ex]
\small
\textsuperscript{1}Dept. of Advanced Imaging, GSAIM, Chung-Ang University, Korea \quad
\textsuperscript{2}Dept. of Artificial Intelligence, Chung-Ang University, Korea \\
\small
\textsuperscript{3}Korea Institute of Industrial Technology (KITECH), Korea \\
{\tt\small \{jincho, cky\}@vilab.cau.ac.kr, choijw@cau.ac.kr} \\[1ex]
}

\begin{document}
\twocolumn[{
    \maketitle
    \vspace{-5mm}
    \centering
    \captionsetup{type=figure}
    \begin{subfigure}{0.19\linewidth}
        \includegraphics[width=\linewidth, height=6.5cm, keepaspectratio]{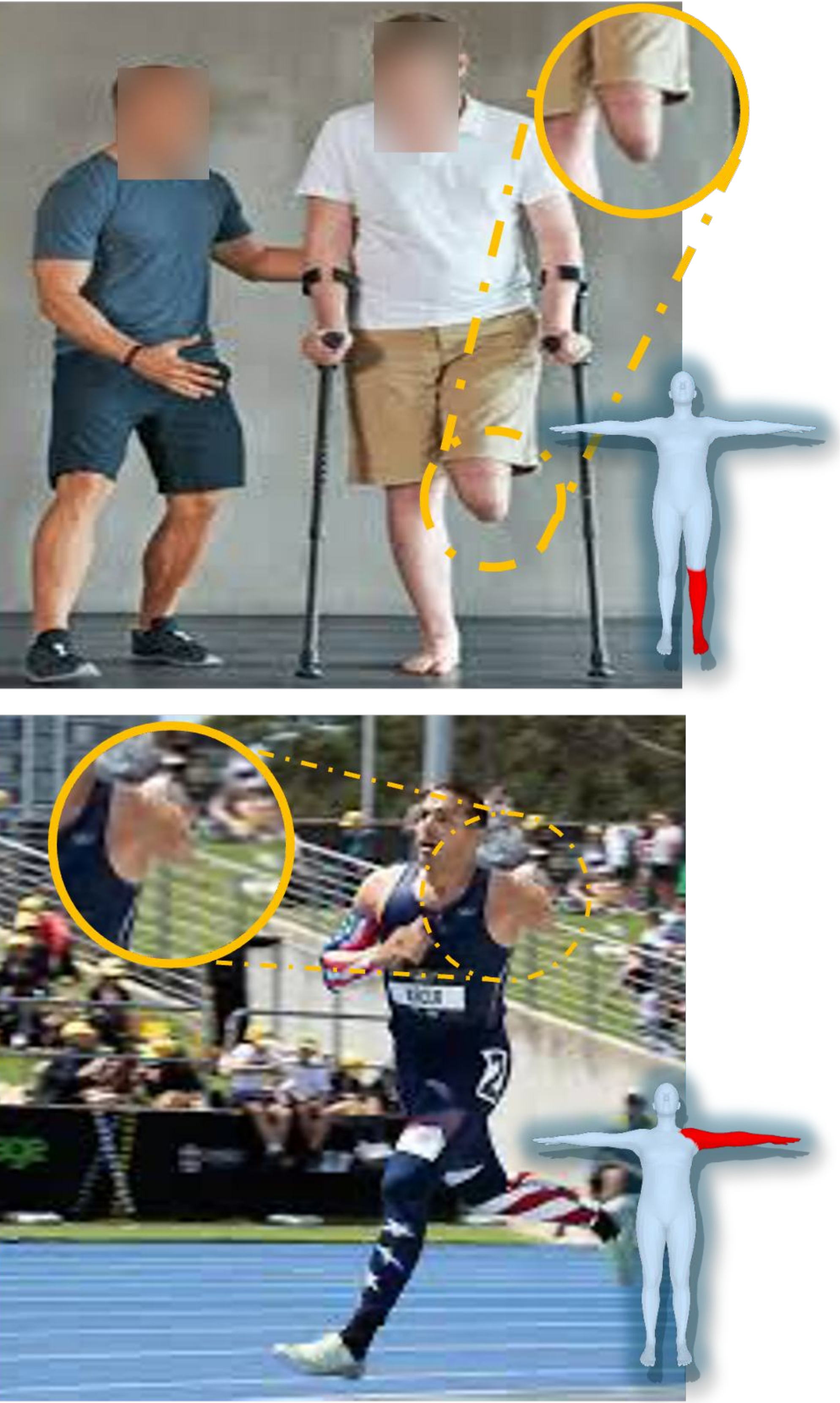} \captionsetup{justification=raggedright,singlelinecheck=false,margin=15pt}
        \caption{Input Image}
    \end{subfigure}
    \hfill
    \begin{subfigure}{0.19\linewidth}
        \includegraphics[width=\linewidth, height=6.5cm, keepaspectratio]{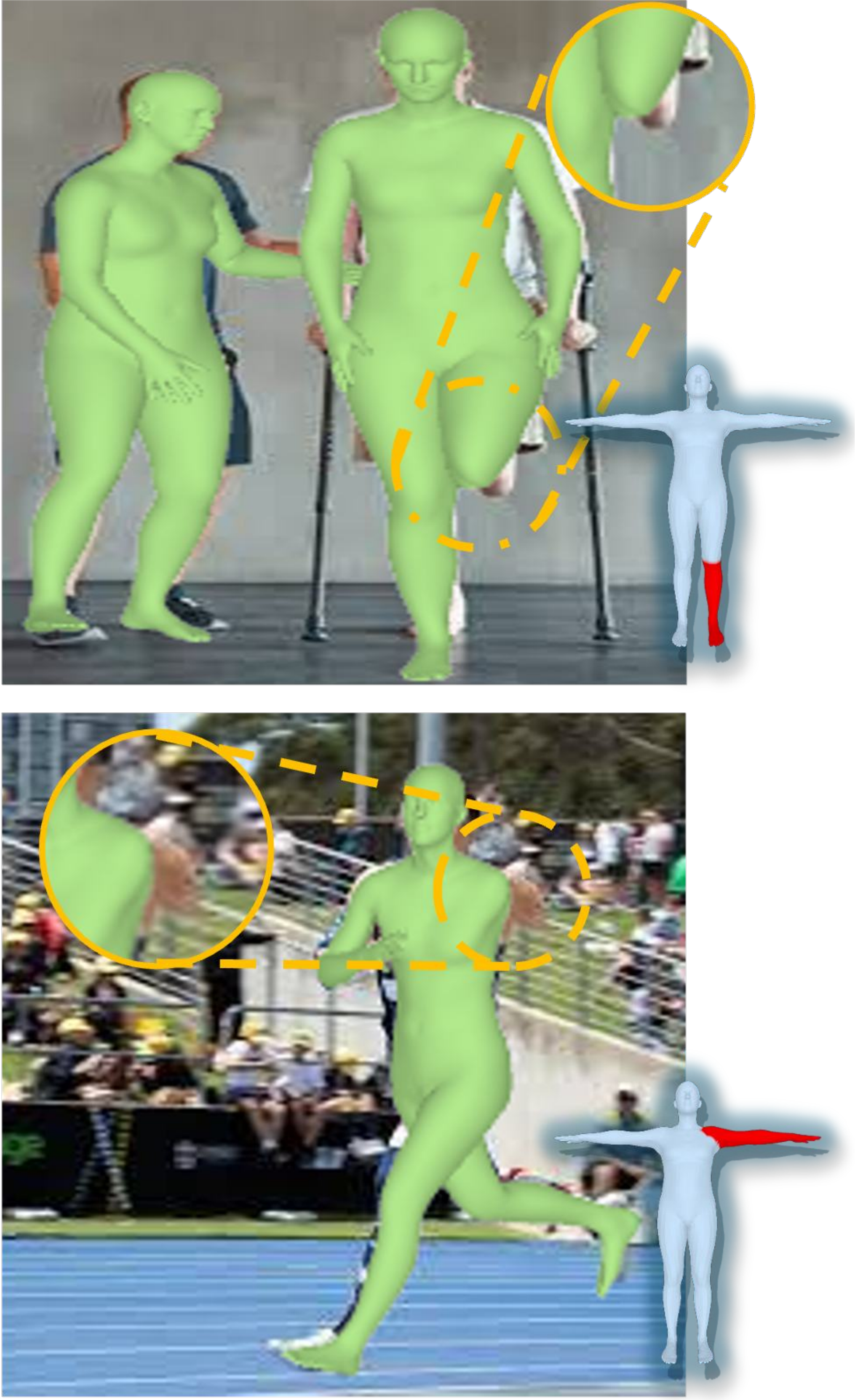}
        \captionsetup{justification=raggedright,singlelinecheck=false,margin=12pt}
        \caption{AJAHR~(Ours)}
    \end{subfigure}
    \hfill
    \begin{subfigure}{0.19\linewidth}
        \includegraphics[width=\linewidth, height=6.5cm, keepaspectratio]{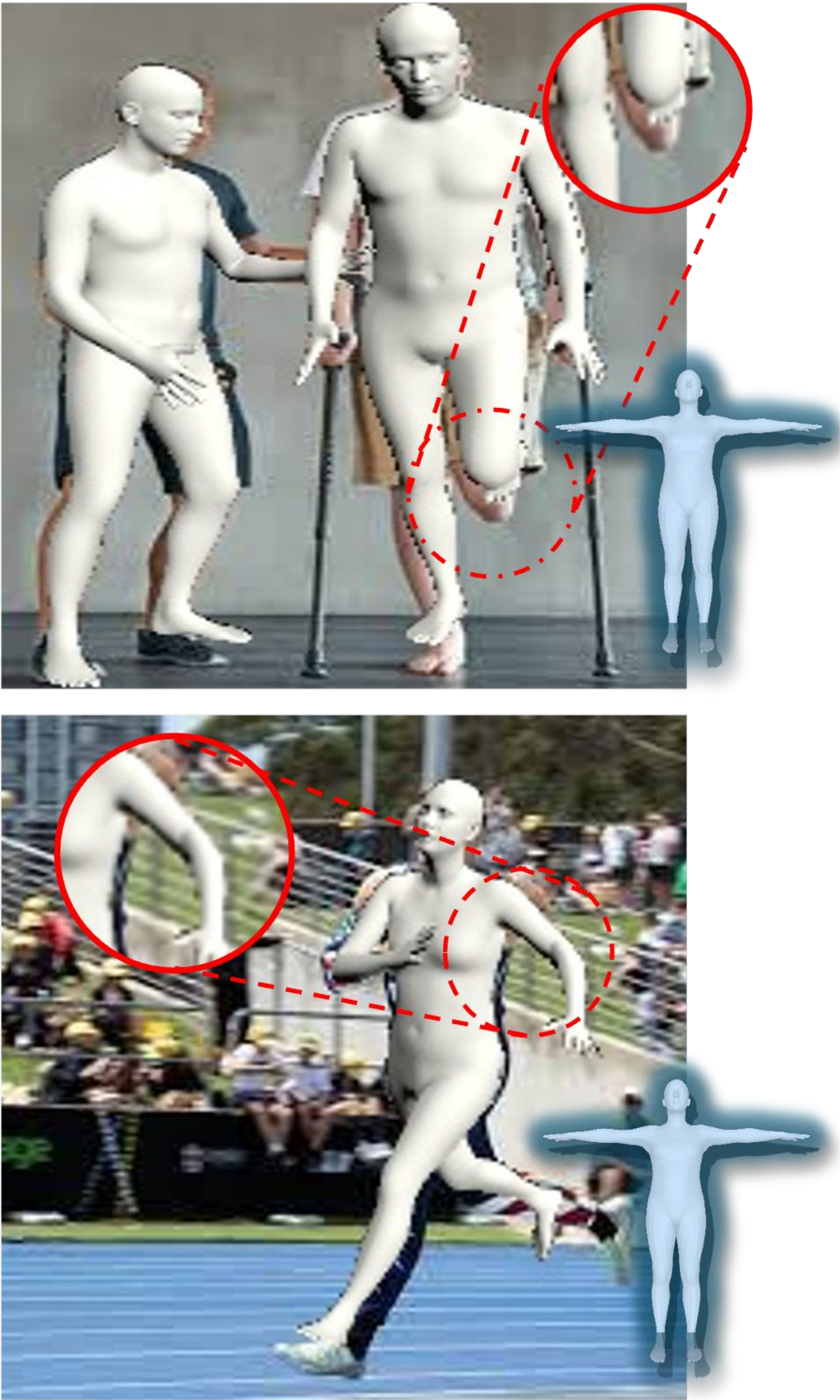}
        \captionsetup{justification=raggedright,singlelinecheck=false,margin=0pt}
        \caption{BEDLAM-CLIFF~\cite{cliff}}
    \end{subfigure}
    \hfill
    \begin{subfigure}{0.19\linewidth}
        \includegraphics[width=\linewidth, height=6.5cm, keepaspectratio]{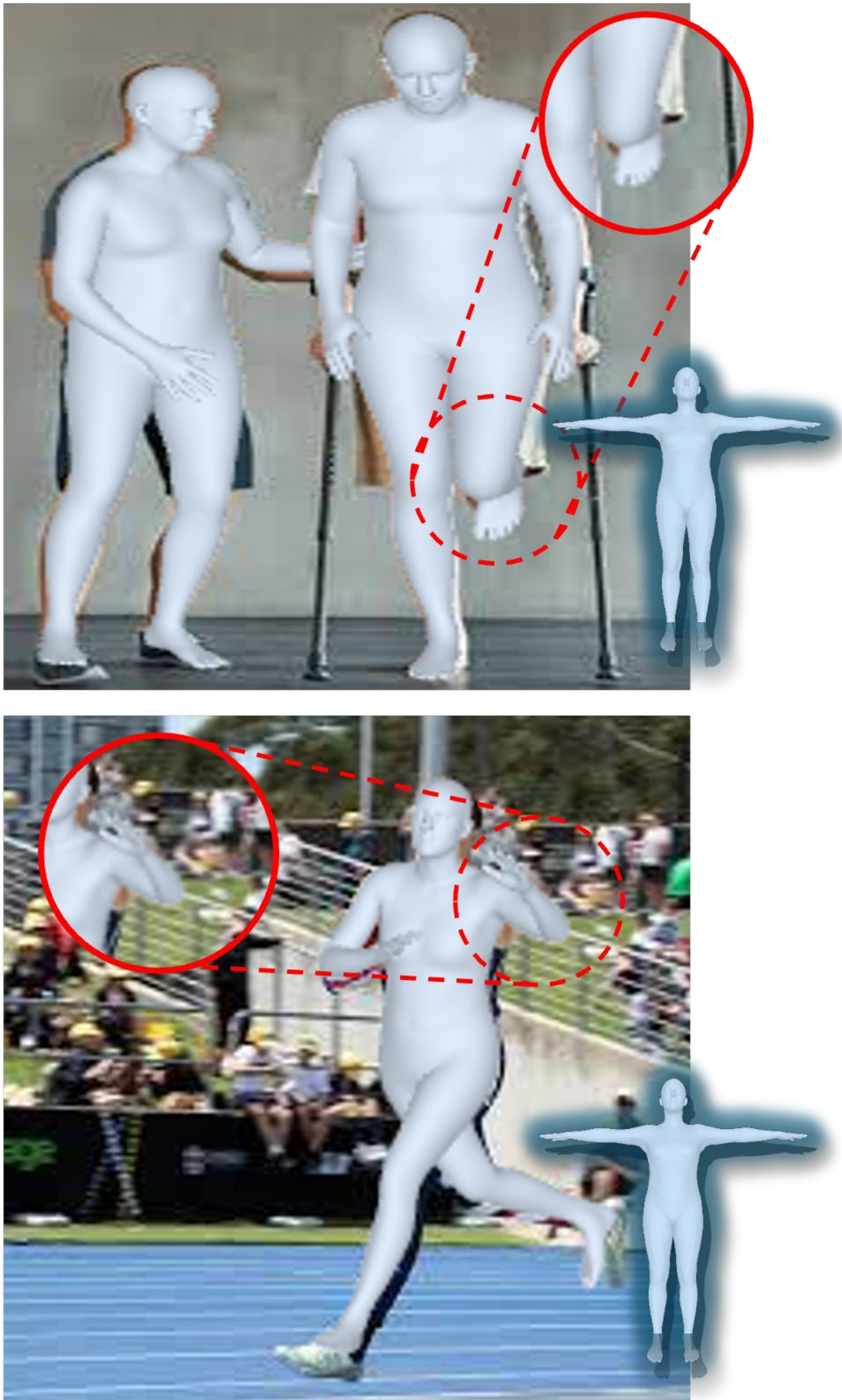}
        \captionsetup{justification=raggedright,singlelinecheck=false,margin=12pt}
        \caption{HMR2.0~\cite{4dhumans}}
    \end{subfigure}
    \hfill
    \begin{subfigure}{0.19\linewidth}
        \includegraphics[width=\linewidth, height=6.5cm, keepaspectratio]{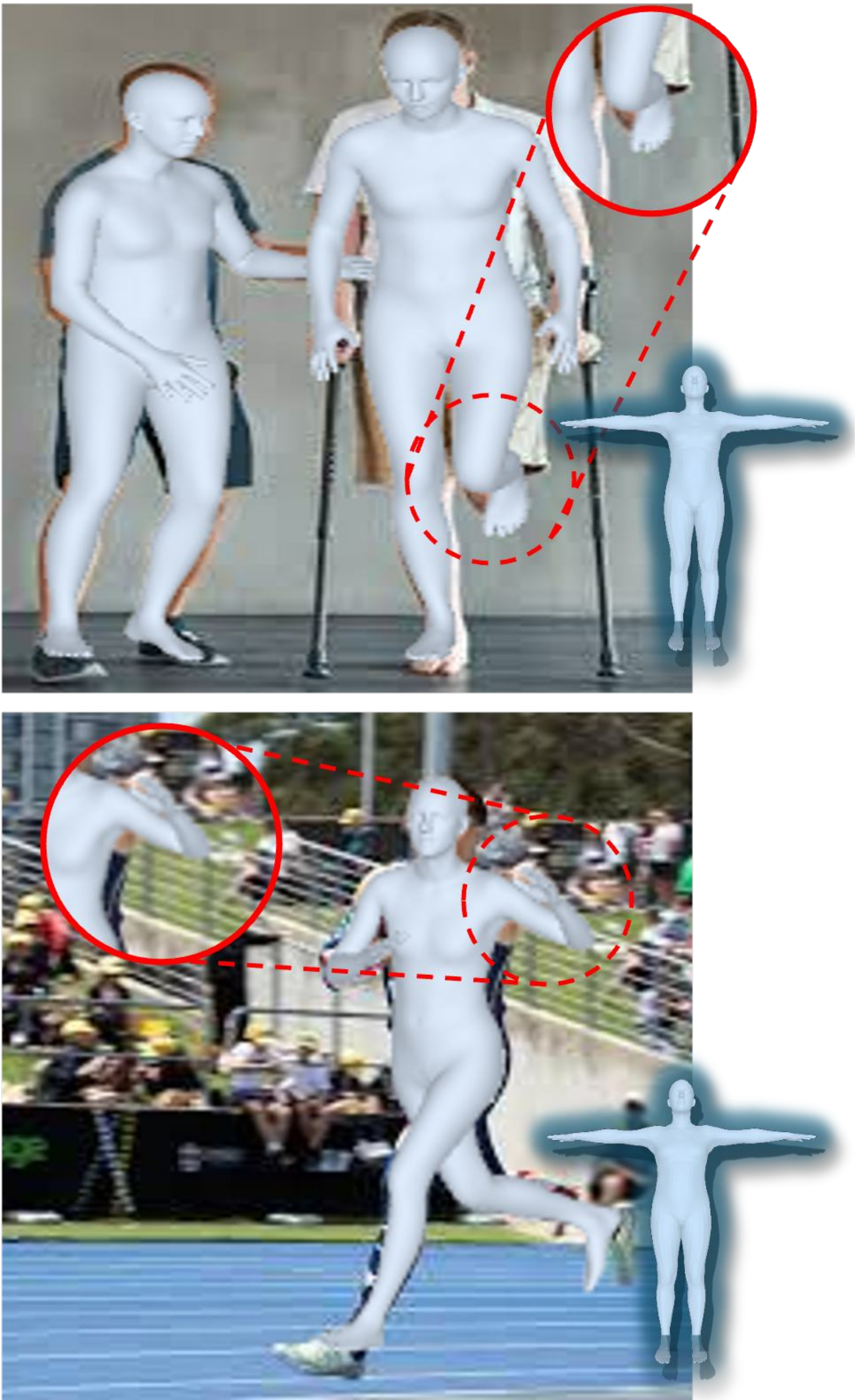}
        \captionsetup{justification=raggedright,singlelinecheck=false,margin=10pt}
        \caption{TokenHMR~\cite{tokenhmr}}
    \end{subfigure}
    \vspace{-7pt}
    \caption{
        \textbf{Examples of Human Mesh Recovery for Amputee and Non-Amputee Individuals.} Column (a) shows input images: the top row includes a non-amputee (left) and an amputee (right), while the bottom row shows an amputee. Our method AJAHR (b) accurately handles both cases, whereas (c), (d), and (e) often misinterpret amputated limbs as intact and infer implausible poses in missing regions.}
    \label{fig:teaser}
    \vspace{5mm}
    }]

\blfootnote{* Equal contribution. \quad \textsuperscript{\dag}Corresponding author.}

\input{sec/0_abstract}
\input{sec/1_intro}

\input{sec/2_related_work}

\input{sec/3_proposed_methods}
\input{sec/4_experiments_copy}
\input{sec/6_conclusion}
\input{sec/supple}

{
    \small
    \bibliographystyle{ieeenat_fullname}
    \bibliography{main}
}

\end{document}

%% file: sec/0_abstract.tex
\vspace{-4.6mm}
\begin{abstract}
Existing human mesh recovery methods assume a standard human body structure, overlooking diverse anatomical conditions such as limb loss. This assumption introduces bias when applied to individuals with amputations—a limitation further exacerbated by the scarcity of suitable datasets. To address this gap, we propose \textbf{A}mputated \textbf{J}oint \textbf{A}ware 3D \textbf{H}uman Mesh \textbf{R}ecovery (\textbf{AJAHR}), which is an adaptive pose estimation framework that improves mesh reconstruction for individuals with limb loss. Our model integrates a body-part amputation classifier, jointly trained with the mesh recovery network, to detect potential amputations. We also introduce Amputee 3D (A3D), which is a synthetic dataset offering a wide range of amputee poses for robust training. While maintaining competitive performance on non-amputees, our approach achieves state-of-the-art results for amputated individuals. Additional materials can be found at: \url{https://chojinie.github.io/project_AJAHR/}
\end{abstract}

%% file: sec/1_intro.tex
\section{Introduction}
\label{sec:intro}
Human pose information is essential for understanding human behavior and enabling effective human-computer interaction. Its importance is evident in applications such as sports~\cite{sports1, sports2}, AR/VR~\cite{vr1} and surveillance~\cite{surveillance}. In computer vision, Human Mesh Recovery (HMR) from a single RGB image offers a cost-effective solution without requiring additional sensors. Although this approach is simple, it has demonstrated strong performance, driving continued research in the field. Our study follows this direction.

Despite recent advancements, a significant limitation remains: most existing methods are trained on datasets composed exclusively of non-amputees, implicitly assuming a standard human body structure. As a result, they tend to produce biased pose estimations when applied to anatomically diverse conditions—such as limb loss—often hallucinating unrealistic poses or body shapes for the missing limbs, rather than accurately reflecting the amputated regions. This highlights the need for more inclusive HMR models that generalize across a wider range of body types.

Prior work has explored model inclusivity by examining the lack of representation in human-related motion datasets and addressing associated bias in pose estimation. Olugbade \etal~\cite{olugbade} highlighted that among 704 publicly available datasets, none include individuals with disabilities performing activities such as sports or daily tasks. In parallel, studies such as Zhou \etal~\cite{prosdif} and WheelPose~\cite{wheelpose} have identified biases in 2D human pose estimation models trained on general human movement datasets.

Inspired by recent findings and limitations, we present AJAHR—an adaptive framework improving mesh recovery in individuals with limb amputations—an area largely overlooked in prior HMR research. Addressing this scenario presents unique challenges: training data with real amputees is extremely scarce, making it difficult to learn body configurations that deviate from standard anatomy. As noted by~\cite{olugbade}, collecting such data—whether in controlled studio environments or in the wild—raises significant ethical and logistical concerns, including safety risks, privacy issues, and high acquisition costs. Moreover, amputation-induced joint absence can be easily confused with occlusions. In the latter case, the joint exists but is merely hidden from view—posing ambiguity for models relying solely on image cues. AJAHR integrates a body-part amputation classifier that is jointly trained with the mesh recovery network, allowing the model to distinguish between amputees and non-amputees and produce pose estimates tailored to each body condition.

As shown in~\cref{fig:teaser}, existing methods often hallucinate unrealistic body parts or fail to represent amputated poses accurately, revealing their limited generalization capability. To the best of our knowledge, no prior dataset or framework has been specifically designed to address this scenario. To fill this gap, we construct Amputee 3D (A3D), a synthetic dataset generated through a controlled data pipeline that offers diverse amputee pose samples for training. Additionally, we compile ITW-amputee, a real-world evaluation set consisting of in-the-wild images of individuals with limb loss collected from online sources. This dataset serves as a benchmark to assess generalization performance on real-world amputee cases, making our contributions a crucial step toward inclusive HMR.
 
Our contributions can be summarized as follows:\\[-0.5em]
\begin{enumerate}[leftmargin=2em]
    \item We introduce and address the first-ever human mesh recovery problem for amputated individuals.
    \item We propose AJAHR, an adaptive HMR model that detects amputated individuals while ensuring stable pose estimation for both amputees and non-amputees.
    \item We construct a new dataset, A3D, for human mesh recovery of amputated individuals and introduce a method to synthesize datasets tailored for this task.
    \item Our approach preserves competitive performance on non-amputees while achieving state-of-the-art results on amputee datasets.
\end{enumerate}

%% file: sec/2_related_work.tex
\section{Related Work}
\label{sec:related}
\noindent\textbf{Inclusiveness in Human Mesh Recovery.} Human mesh recovery research has primarily focused on improving pose accuracy in occluded conditions~\cite{3doh, 3dnbf, pare} or leveraging motion capture datasets like AMASS~\cite{AMASS} and MOYO~\cite{moyo} to enhance pose priors across various postures~\cite{smplify, hmr, tokenhmr}. While these methods improve accuracy, they largely focus on individuals with typical anatomical structures, overlooking anatomical diversity. To improve inclusivity, some studies have specifically explored individuals with physical disabilities~\cite{wheelpose, prosdif}. For instance, WheelPose~\cite{wheelpose} introduced a synthesis pipeline for wheelchair users, while Zhou \etal~\cite{prosdif} reconstructed prosthetic limbs as intact limbs to enable robust pose estimation. However, human pose estimation or human mesh recovery for individuals with limb amputations remains unexplored. To address this gap, we introduce a synthesis pipeline for amputee pose data and analyze model adaptability to missing body parts.

\vspace{1mm}
\noindent\textbf{Monocular 3D Human Mesh Recovery.}
3D human mesh recovery from a single RGB image involves extracting visual features to reconstruct a parametric human body. Existing methods are broadly categorized into regression- and optimization-based approaches. Regression-based methods~\cite{hmr, 4dhumans, pare, cliff, indirect, self} directly predict body model parameters~\cite{SMPL, ghum, STAR} in a single forward pass, enabling real-time inference. Optimization-based methods~\cite{smplify, smplx, spin, scorehmr} estimate these parameters by fitting the SMPL model to 2D cues such as keypoints and silhouettes, and iteratively refine predictions using additional image information. Despite strong performance, most methods are trained on datasets without disabled individuals, resulting in pose priors that generalize poorly to amputees. To address this, we adopt TokenHMR~\cite{tokenhmr} as our baseline, which reframes pose estimation as a token classification task to mitigate bias and incorporate structured priors. Building on this framework, we integrate pose priors for individuals with limb amputations, improving prediction accuracy for missing body regions.

%% file: sec/3_proposed_methods.tex
\begin{figure*}[t]
    \centering
    \includegraphics[width=\textwidth]{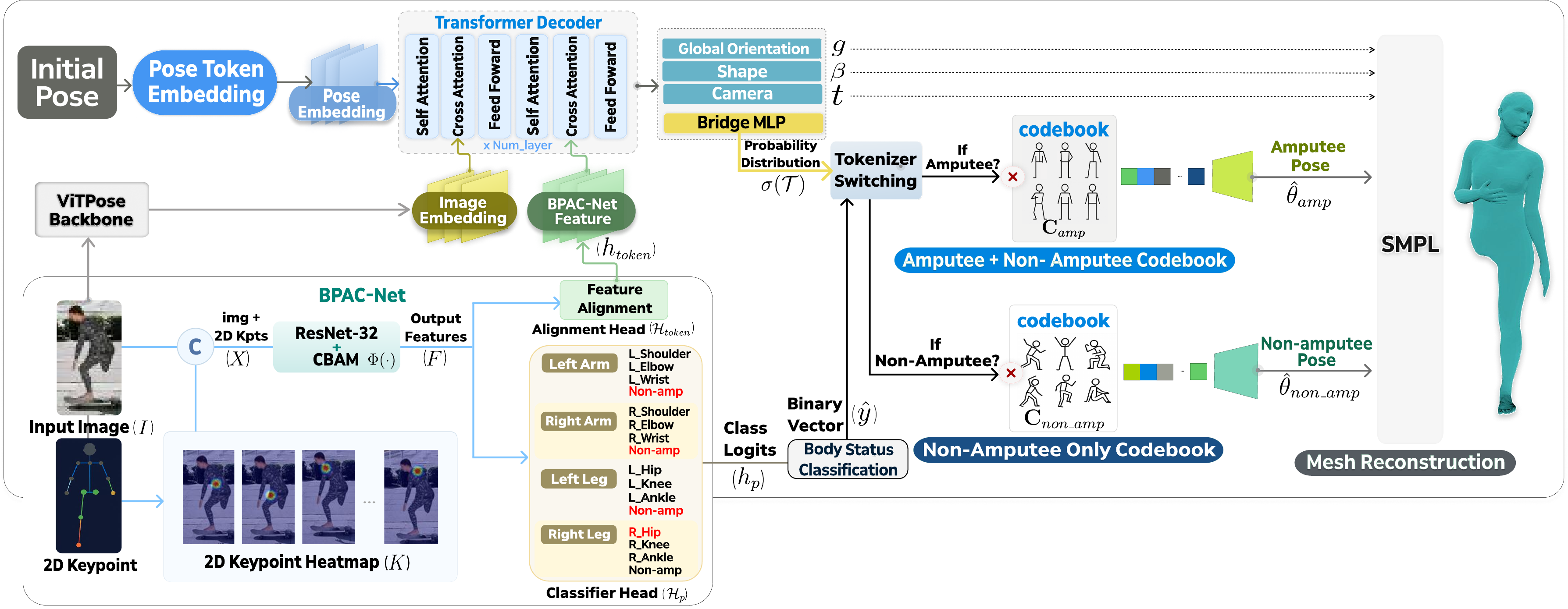}
    \caption{~\textbf{Overview of AJAHR Architecture.} The proposed model takes typical human pose datasets and the A3D dataset—a synthetic dataset of amputated individuals (see~\cref{sec:synthesis_pipeline})—as input. It employs BPAC-Net, a body-part amputation classifier (see~\cref{sec:BPAC-Net}), to detect limb absence and uses this information to guide mesh recovery for both amputees and non-amputees. Based on the predicted body-part status, the model connects to different pre-trained tokenizers (see~\cref{sec:Tokenizer}), which remain frozen during training while the rest of the model is optimized accordingly.}
    \label{fig:overall_framework}
    \vspace{-2mm}
\end{figure*}

\section{Proposed Method}
\subsection{Preliminaries} \label{sec:preliminary}
\noindent\textbf{Body Model.} The SMPL~\cite{SMPL} is a low-dimensional, differentiable parametric body model that represents the human body. The model takes as an input the pose parameters $\theta \in \mathbb{R}^{24 \times 3 \times 3}$ and shape parameters $\beta \in \mathbb{R}^{10}$. As an output, the model generates the human body mesh, $\mathcal{M}$, which consists of $V \in \mathbb{R}^{N \times 3}$, where $N$=6890 represents the number of vertices. The 3D joints, \(J_{3D}\), are obtained by combining the pre-trained joint regressor with the vertices.

\vspace{1mm}
\noindent\textbf{Representation of Amputation under the SMPL.} \label{sec:representation_amputation_smpl} The SMPL model has a kinematic tree structure, where 24 joints are organized in a parent-child relationship. As it is pre-trained and not directly trainable, we leverage this structure to represent amputations without modifying the model.
Our method encodes amputations by setting the pose parameter $\theta$ of the amputated parent joint and all its descendants to a zero matrix. When these parameters are passed into the SMPL model, the corresponding vertices collapse into a single location near the amputated joint, effectively simulating limb absence. The generated vertices are then multiplied by a pre-trained joint regressor to obtain 3D joint positions, resulting in the child joints of the amputated region shifting toward the amputated parent due to the hierarchical structure of the body model. 
For details, please refer to Supplementary Sec. B.

\begin{table}[t]
\input{table/dataset_comparison}
\label{dataset_comp}
\end{table}

\subsection{Amputee Dataset Synthesis} \label{sec:synthesis_pipeline}
\noindent\textbf{Overview.} Training a pose estimator for individuals with limb amputations requires annotated amputee images. However, collecting such data is costly and raises concerns regarding safety, accessibility, and diversity. In addition, state-of-the-art generative models (e.g., GPT-4~\cite{chatgpt}, ControlNet~\cite{controlnet}) still fall short in producing sufficient high-quality amputee images for our purposes. To address these limitations, we synthesize amputee data through a multi-stage pipeline that generates diverse amputee instances with various ethnic appearances and realistic backgrounds. This pipeline includes an index selection module to simulate amputated body parts in the SMPL representation, and a controller module that handles texture assignment for gender and ethnic diversity. See Supplementary Sec. A for architectural design and implementation details.

\vspace{1mm}
\noindent\textbf{Synthesizing Amputee Representations.} We leverage off-the-shelf models designed for recovering human mesh from a single RGB image using the SMPL body model to create synthetic representations of individuals with limb amputations, where amputated parts are represented using a zero pose. Additionally, we enhance visual diversity by incorporating skin and clothing textures from BEDLAM~\cite{bedlam} assets, which reflect a range of human appearances and contribute to comprehensive representation.

\vspace{1mm}
\noindent\textbf{Generating Background Images.} 
To ensure generalization, we utilized both indoor and in-the-wild pose images from diverse environments. To obtain clean background images, we applied a two-stage removal process to benchmark datasets~\cite{h36m_pami, mpii, mscoco}: a segmentation model~\cite{SAM} was used to detect and mask human regions, which were then inpainted using LaMa~\cite{lama}, a model designed for object removal. 
This process generated approximately 3K backgrounds from MSCOCO~\cite{mscoco} and 1K from MPII~\cite{mpii}. For Human3.6M~\cite{h36m_pami}, which consists of scenes captured from four fixed camera viewpoints in a controlled multi-view environment, we extracted one background image per viewpoint, resulting in a total of four. Finally, we projected the synthesized amputee meshes onto these cleared backgrounds using weak perspective projection.

\noindent\textbf{A3D: Amputee 3D Pose Dataset.}  
Through our synthetic data pipeline, we generated over one million high-quality images simulating various amputation scenarios. These images reflect a diverse range of poses, clothing, and backgrounds, contributing to both generalization and robustness in training. Inspired by BEDLAM~\cite{bedlam}, we designed our dataset to reflect balanced demographic diversity: African, Asian, and Indian each account for 20\% of the mannequin population, while Hispanic, Middle Eastern, Southeast Asian, and White each comprise 10\%. To reflect a wide range of amputation types, we simulated limb loss across multiple body parts—including hand, forearm, full arm, ankle, knee, and full leg—based on poses commonly observed in benchmark datasets of non-amputee subjects. Alongside the images, we provide full annotations, including SMPL parameters, 3D and 2D joint coordinates, and detailed amputation region labels, all aligned with ground truth (GT). As shown in \cref{tab:dataset_comparison} shows that our dataset is sufficiently large and serves as the first dataset tailored for the amputee domain. During training, we leverage 2D keypoints, 3D keypoints, and SMPL parameters as supervision signals. For 2D keypoints, we follow standard practices by setting the coordinates of joints in non-visible regions—whether due to amputation or occlusion (i.e., the missing joint and its child joints)—to (0, 0), thereby excluding them from 2D supervision. In contrast, 3D keypoints and SMPL parameters corresponding to amputated regions are still used as supervision signals, allowing the model to learn representations for structurally absent but semantically defined joints. This distinction between modality-specific supervision handling for occlusion and amputation enables accurate learning based on the SMPL framework without requiring architectural modifications or additional retraining.

\subsection{Architecture of AJAHR}
\noindent\textbf{Overview.} As illustrated in~\cref{fig:overall_framework}, AJAHR adopts a Vision Transformer (ViT) \cite{vit} based architecture inspired by HMR2.0~\cite{4dhumans} and TokenHMR~\cite{tokenhmr}.
Input images are encoded into embedding tokens via a ViT encoder and then refined through a Transformer decoder~\cite{transformer} featuring two cross-attention pathways.
First, a predefined zero‑pose parameter token attends to the image feature tokens to initialize the pose representation.
Second, the classifier-generated token undergoes additional cross-attention within the decoder, allowing semantic cues from the classifier to guide the pose regression.
Additionally, BPAC-Net classifies the amputation status of the body by taking the input image and its corresponding 2D keypoints as input. During training, ground-truth 2D keypoints are used, while at inference time, predicted keypoints from the ViTPose~\cite{vitpose} detector are employed.

The output tokens from the transformer decoder~\cite{transformer} are split into four branches, each corresponding to a major body region—left arm, right arm, left leg, and right leg.  Each branch performs amputation classification by determining whether any child joints within its region, based on the SMPL kinematic tree, are amputated or non-amputated.  
In one branch, a bridge MLP transforms the decoder output to match the dimensionality of the codebook, producing logits for each codebook entry. Applying the softmax function yields a probability distribution that serves as soft weights over the pre-trained codebook, each entry representing a latent pose component. The pre-trained codebook consists of two types: one trained on both amputee and non-amputee datasets, and another trained only on non-amputee data. Based on the amputation status predicted by BPAC-Net, a tokenizer-switching strategy selects the appropriate codebook. The selected codebook is then aggregated via multiplication using the token distribution as weights, resulting in the final predicted pose parameter, $\hat{\theta}$. 
The other three branches independently regress the global rotation ($g$), body shape parameters ($\beta$), and camera translation ($t$) through separate regression heads.
Finally, the predicted parameters are passed to SMPL for mesh reconstruction.

We follow the TokenHMR~\cite{tokenhmr} paradigm by pretraining the tokenizer separately. Specifically, the codebook and decoder are trained in advance, and then frozen during pose estimation training. 

\subsubsection{Body Part Amputation Classifier (BPAC-Net)}
\label{sec:BPAC-Net}
The proposed BPAC-Net serves three key roles:
(1)~\textbf{Loss Adjustment:} Enhancing learning on amputee data by increasing loss for amputated regions.
(2)~\textbf{Implicit Learning Assistance:} BPAC-Net features undergo cross-attention with the transformer decoder~\cite{transformer} to improve pose estimation for amputated parts.
(3)~\textbf{Visualization:} Enforcing zero values in SMPL pose parameters for amputated joints to ensure accurate visual representation.

To address the ambiguity of missing or occluded limbs in RGB images, we incorporate 2D keypoint heatmaps as additional visual cues. For each amputee and non-amputee image, we incorporate its corresponding keypoint heatmap as an additional visual cue by concatenating the RGB input $I \in \mathbb{R}^{H \times W \times 3}$ with the keypoints $K \in \mathbb{R}^{H \times W \times J}$ along the channel dimension to form $X = \mathrm{concat}(I, K)$. These combined inputs are fed into $\Phi(\cdot)$, which integrates ResNet-32~\cite{resnet} and CBAM~\cite{woo2018cbam} to extract spatial and semantical feature maps $F = \Phi{(X)}\in \mathbb{R}^{ h \times w \times c}$.

The extracted features are passed to four parallel classification heads \(\mathcal{H}_p\) and one feature alignment head \(\mathcal{H}_{token}\). Four classification heads, denoted as 
\(\mathcal{H}_p \in \{\mathcal{H}_{L_{arm}}, \mathcal{H}_{R_{arm}}, \mathcal{H}_{L_{leg}}, \mathcal{H}_{R_{leg}}\}\), 
are responsible for predicting the amputation status of each limb. For each body part 
\(p \in \{L_{arm}, R_{arm}, L_{leg}, R_{leg}\}\), 
the corresponding classification head outputs a part-specific logit feature vector 
\(h_p = \mathcal{H}_p(F) \in \mathbb{R}^{4}\). Here, $ h_p \in \mathbb{R}^{4} $ represents the logits for four classes corresponding to each limb, as shown in the head components of BPAC-Net in \cref{fig:overall_framework}.
These predictions are then compared to the ground-truth amputation or non-amputation labels \(lb\)  
using a cross-entropy loss function.

\begin{figure}[t]
    \centering
    \includegraphics[width=1\linewidth]{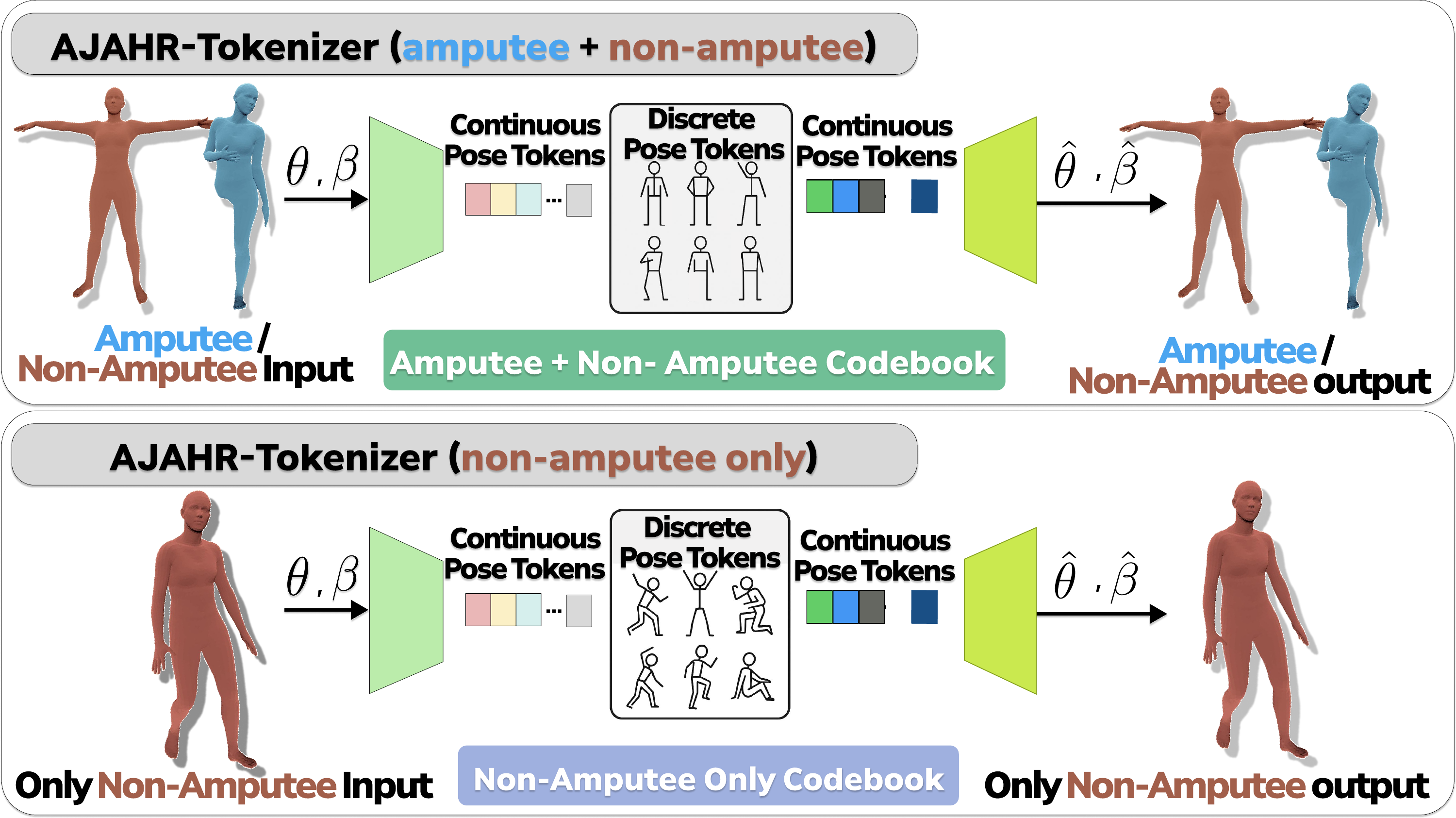}
    \caption{\textbf{Tokenizing for diverse pose prior.} The input \( \theta \) consists of SMPL pose parameters from amputees in the A3D dataset and non-amputees in AMASS~\cite{AMASS} and MOYO~\cite{moyo}. This allows the tokenizer to encode amputee pose information into the codebook, enabling the trained decoder to incorporate pose priors for both amputees and non-amputees, thereby improving its ability to reconstruct diverse human poses.}
    \label{fig:tokenizer}
    \vspace{-2mm}
\end{figure}

To enable the tokenizer switching strategy, we determine whether any of the predicted classes correspond to an amputated state. Each classification head \(\mathcal{H}_p\) outputs a logit vector \(h_p \in \mathbb{R}^4\), 
where class 0 corresponds to a non-amputated limb, and classes 1, 2, and 3 represent 
amputation types specific to the body part that the head is responsible for. 
A binary decision \(\hat{y}_p \in \{0, 1\}\) is obtained as:

\begin{equation}
\hat{y}_p = \begin{cases}
0 & \text{if } \arg\max(h_p) = 0, \\
1 & \text{otherwise.}
\end{cases}
\label{eq:arch1}
\end{equation}

The values obtained for the four parts are concatenated to form the 4-dimensional vector $\hat{{y}}=\bigl[\hat{y}_{L_{arm}},\,
      \hat{y}_{R_{arm}},\,
      \hat{y}_{L_{leg}},\,
      \hat{y}_{R_{leg}}\bigr]^T$, representing the predicted amputation status for each limb.

The feature alignment head, denoted as $\mathcal{H}_{\text{token}}$, produces a global feature vector $\mathcal{H}_{\text{token}}(F) \in \mathbb{R}^{1280}$,  
which is directly used as the cross-attention mechanism of the transformer decoder~\cite{transformer}. 
The BPAC-Net classification loss is defined as a sum over four limb-specific predictions:
\vspace{-3pt}
\begin{equation} \label{eq:bpac_loss}
\mathcal{L}_{cls} = \sum_{p \in \{L_{arm}, R_{arm}, L_{leg}, R_{leg}\}} CE(h_p, lb),
\end{equation}
where $CE$ denotes the cross-entropy loss function. Details of the amputation labels can be found in the BPAC-Net head outputs in~\cref{fig:overall_framework} and the Supplementary Sec. G.


\subsubsection{AJAHR-Tokenizer}
\label{sec:Tokenizer}
\noindent\textbf{Tokenizer Switching Strategy.}
In this study, we present two versions of the tokenizer as depicted in \cref{fig:tokenizer}: one trained on a combined dataset of amputee and non‑amputee poses, and the other trained exclusively on non‑amputee poses. Once each tokenizer has been trained, the codebook \(\mathbf{C}_{amp}\), which can reconstruct both amputee and non-amputee poses, and the codebook \(\mathbf{C}_{non\_amp}\), specialized for non-amputee poses, along with their corresponding decoders, are kept frozen. Then, following the tokenizer switching strategy, the predicted binary indicator \(\hat{\mathbf{y}}\) from BPAC-Net is applied to the output tokens of the transformer decoder to infer the pose parameters as follows:
\begin{equation} \label{eq:arch}
\begin{gathered}
\hat{\theta} =
\begin{cases} 
\sigma{(\mathcal{T})}\times\mathbf{C}_{amp} & \text{if } \|\hat{y}\|_1 > 0, \\
\sigma{(\mathcal{T})}\times\mathbf{C}_{non\_amp} & \text{otherwise,}
\end{cases}
\end{gathered}
\end{equation}
where \(\hat{\theta}\) is composed of \(\hat{\theta}_{amp}\), obtained from \(\mathbf{C}_{amp}\), and \(\hat{\theta}_{non\_amp}\), obtained from \(\mathbf{C}_{non\_amp}\), and is used for final mesh reconstruction.

\begin{table*}[t]
\centering
\begin{minipage}[t]{0.48\textwidth}
\centering
\resizebox{\columnwidth}{!}{
\begin{tabular}{c|ccc|ccc}
\hline
\multirow{2}{*}{Method} & \multicolumn{3}{c|}{A3D} & \multicolumn{3}{c}{ITW-amputee} \\ 
                        & MVE$\downarrow$ & MPJPE$\downarrow$ & PA-MPJPE$\downarrow$
                        & MVE$\downarrow$ & MPJPE$\downarrow$ & PA-MPJPE$\downarrow$ \\
\hline
HMR2.0~\cite{4dhumans}       & 89.35 & 96.75 & 86.14 & \textbf{110.33} & 154.43 & 121.83 \\
BEDLAM-CLIFF~\cite{bedlam, cliff} & 83.38 & 88.12 & 56.45 & 128.09 & 150.12 & 117.74 \\
TokenHMR~\cite{tokenhmr}     & 76.01 & 74.70 & 49.94 & 136.52 & 146.12 & 91.00 \\
AJAHR (Ours)             & \textbf{73.42} & \textbf{73.19} & \textbf{49.42} & 116.42 & \textbf{129.25} & \textbf{77.18} \\
\hline
\end{tabular}
}
\vspace{-2mm}
\caption{\textbf{Results on Amputee Data.}}
\label{tab:amputation_table}
\end{minipage}%
\hfill
\begin{minipage}[t]{0.48\textwidth}
\centering
\resizebox{\columnwidth}{!}{
\begin{tabular}{c|ccc|ccc}
\hline
\multirow{2}{*}{Method} & \multicolumn{3}{c|}{EMDB~\cite{emdb}} & \multicolumn{3}{c}{3DPW~\cite{3dpw}} \\
                        & MVE$\downarrow$ & MPJPE$\downarrow$ & PA-MPJPE$\downarrow$
                        & MVE$\downarrow$ & MPJPE$\downarrow$ & PA-MPJPE$\downarrow$ \\
\hline
HMR2.0~\cite{4dhumans}        & 141.41 & 117.66 & 75.89 & 95.29 & 81.64 & 53.95 \\
BEDLAM-CLIFF~\cite{bedlam, cliff} & 129.00 & 97.88  & 62.40 & 99.32 & 76.45 & 51.21 \\
TokenHMR~\cite{tokenhmr}     & 113.26 & 93.77  & 58.98 & \textbf{90.23} & 72.87 & 47.17 \\
AJAHR (Ours)             & \textbf{112.83} & \textbf{91.74} & \textbf{58.62} & 95.26 & \textbf{71.77} & \textbf{44.94} \\
\hline
\end{tabular}
}
\vspace{-2mm}
\caption{\textbf{Results on Non-Amputee Data.}}
\label{tab:nonamputation_table}
\end{minipage}
\end{table*}

\input{table/rebuttal/rebuttal_combined}

\noindent\textbf{Training Objective for AJAHR-Tokenizer.}
\label{sec:tokenizer_training_objective} As described in \cref{sec:representation_amputation_smpl}, the AJAHR-Tokenizer predicts full-body pose parameters \(\theta\) to reconstruct both amputee and non-amputee poses. Directly assigning zero values to amputated joints in 3D rotation representations can induce numerical instability (e.g., NaNs). To mitigate this, original pose parameters are preserved during training, and zero masking is applied in post-processing according to the predicted body-part status \(\hat{y}_p\) from BPAC-Net, where \(\hat{y}_p \in \{0,1\}^4\) is a binary indicator vector over \(4\) predefined body parts (1 denotes an amputated part).

The AJAHR-Tokenizer is trained on a mix of large-scale non-amputee pose data (AMASS~\cite{AMASS}, MOYO~\cite{moyo}) and amputee-specific SMPL poses from the A3D dataset. To ensure stable 3D rotation representation, each joint is encoded using the continuous 6D representation from~\cite{zhou2019continuity}.

The process encodes the pose parameters via an encoder \(\mathbf{E}\) into latent features \(Z = \mathbf{E}(\theta) = [z_1, z_2, \ldots, z_S]\), where \(z_i \in \mathbb{R}^{d}\) and \(S\) is the number of tokens. A learnable codebook \(\mathbf{C} = \{c_m\}_{m=1}^M\) is maintained, with each code \(c_m \in \mathbb{R}^{d}\) and \(d\) denoting the dimensionality of the code vectors. Each latent vector \(z_i\) is quantized to its nearest codebook entry:

\begin{equation}
\tilde{z}_i = \arg\min_{c_m \in \mathbf{C}} \| z_i - c_m \|_2,
\end{equation}
producing the quantized set \(\tilde{Z} = [\tilde{z}_1, \tilde{z}_2, \ldots, \tilde{z}_S]\).

Following VQ-VAE~\cite{vqvae} and TokenHMR~\cite{tokenhmr}, the total loss combines three components: mixed reconstruction loss \(\mathcal{L}_{\text{mix}}\), codebook (embedding) loss, and commitment loss. 
\begin{equation}
\mathcal{L}_{\text{total}}
= \lambda_{\text{mix}} \mathcal{L}_{\text{mix}}
+ \lambda_{\text{cb}} \| \mathrm{sg}[Z] - \tilde{Z} \|_2^2
+ \lambda_{\text{com}} \| Z - \mathrm{sg}[\tilde{Z}] \|_2^2,
\label{eq:tokenizer_loss}
\end{equation}
where \(\mathrm{sg}[\cdot]\) is the stop-gradient operator. The mixed loss \(\mathcal{L}_{\text{mix}}\) measures \(\ell_2\) distance between predicted and ground-truth mesh vertices \(V\), 3D joints \(J_{3D}\), and pose parameters \(\theta\).  The loss weights are set to $\lambda_{\text{mix}} = 100.0$, $\lambda_{\text{cb}} = 1.0$, and $\lambda_{\text{com}} = 1.0$. To prevent codebook collapse, we apply codebook reset and exponential moving average (EMA) updates as in prior work~\cite{t2m-gpt, tokenhmr}. Full training details are provided in Supplementary Sec.~H.

\vspace{-4mm}
\subsubsection{AJAHR Losses}
The overall loss of AJAHR combines SMPL-related regression terms with the amputation-aware classification loss from BPAC-Net.  
Let \(\theta, \beta\) and \(\hat{\theta}, \hat{\beta}\) be the ground-truth and predicted SMPL pose and shape parameters, respectively.  
Predicted 3D joint locations \(\hat{J}_{3D}\) are obtained via a pretrained joint regressor from \(\hat{\theta}, \hat{\beta}\), and projected to 2D as \(\hat{J}_{2D}\), with ground-truth targets \(J_{3D}, J_{2D}\).  
The pose loss \(\mathcal{L}_\theta(\theta, \hat{\theta})\) measures the \(\ell_2\) distance in a stable (e.g., 6D) rotation representation, and \(\mathcal{L}_\beta(\beta, \hat{\beta})\) is the \(\ell_2\) distance between shape coefficients.  
These are complemented by the 3D joint loss \(\mathcal{L}_{3D}(J_{3D}, \hat{J}_{3D})\), 2D projection loss \(\mathcal{L}_{2D}(J_{2D}, \hat{J}_{2D})\), and BPAC-Net’s classification loss \(\mathcal{L}_{\text{cls}}\), all computed as \(\ell_2\) or cross-entropy losses.  
The overall loss \(\mathcal{L}_{overall}\) is defined as:
\vspace{-5mm}
\input{equtations/total_loss}
\noindent We set the hyperparameter as follows: \(\lambda_{\theta}=10^{-3}\), \(\lambda_{\beta}=5\cdot 10^{-4}\), \(\lambda_{3D}=5\cdot 10^{-2}\), \(\lambda_{2D}=10^{-2}\), and \(\lambda_{\text{cls}}=10^{-2}\).

%% file: table/dataset_comparison.tex
\centering
\large 
\renewcommand{\arraystretch}{1.15} 
\resizebox{\linewidth}{!}{ 
    \begin{tabular}{l c c c c c c}
        \toprule
        \textbf{Dataset} & \textbf{Year} & \textbf{Type} & \textbf{Annot.} & \textbf{SMPL GT} & \textbf{Amputee} & \textbf{\# Images} \\
        \midrule
        WheelPose~\cite{wheelpose} & 2024 & Image & 2D & \texttimes & \texttimes & - \\
        BEDLAM~\cite{bedlam}$^{*}$ & 2023 & Video & 3D+2D & \checkmark & \texttimes  & 18M \\
        EMDB~\cite{emdb}$^{*}$ & 2023 & Video & 3D+2D & \checkmark & \texttimes & 105K \\
        3DPW~\cite{3dpw}$^{*}$ & 2018 & Video & 3D+2D & \checkmark & \texttimes & 53K \\
        Human3.6M~\cite{h36m_pami}$^{*}$ & 2015 & Video & 3D+2D & \texttimes & \texttimes & 3.6M \\ 
        MSCOCO~\cite{mscoco} & 2014 & Image & 2D & \texttimes & \texttimes & 200K \\
        MPII~\cite{mpii} & 2014 & Image & 2D & \texttimes & \texttimes & 25K \\
        \textbf{Ours (A3D)} & 2025 & Image & 3D+2D & \checkmark & \checkmark & 1.0M \\
        \bottomrule
    \end{tabular}
}
\caption{\textbf{Comparison of datasets.} $^*$ indicates that the number of images refers to frames extracted from videos.}
\label{tab:dataset_comparison}

%% file: table/rebuttal/rebuttal_combined.tex
\begin{table*}[!t]
  \centering
  \large 
  \setlength{\tabcolsep}{1.3pt} 
  \renewcommand{\arraystretch}{1.00} 
  \resizebox{\linewidth}{!}{
  \begin{tabular}{llc|ccc|ccc|ccc|ccc}
    \hline
    & \textbf{Experiments} & \textbf{Use Classifier}
      & \multicolumn{3}{c|}{\textbf{EMDB}~\cite{emdb}} 
      & \multicolumn{3}{c|}{\textbf{3DPW}~\cite{3dpw}} 
      & \multicolumn{3}{c|}{A3D} 
      & \multicolumn{3}{c}{ITW-amputee} \\
    & & & MVE$\downarrow$ & MPJPE$\downarrow$ & PA-MPJPE$\downarrow$
      & MVE$\downarrow$ & MPJPE$\downarrow$ & PA-MPJPE$\downarrow$
      & MVE$\downarrow$ & MPJPE$\downarrow$ & PA-MPJPE$\downarrow$
      & MVE$\downarrow$ & MPJPE$\downarrow$ & PA-MPJPE$\downarrow$ \\
    \hline

(a) & Noise Ratio : 100\% & \checkmark
    & 117.71 & 96.22 & 60.97 & 99.03 & 75.64 & 49.31 & 91.30 & 91.21 & 71.31 & 144.08 & 147.41 & 88.08 \\
    & Noise Ratio : 75\% & \checkmark
    & 115.77 & 94.78 & 59.31 & 97.91 & 73.31 & 46.88 & 89.12 & 89.32 & 69.74 & 142.21 & 145.99 & 86.51 \\
    & Noise Ratio : 50\% & \checkmark
    & 115.31 & 94.12 & 59.22 & 97.43 & 72.77 & 45.87 & 88.76 & 88.98 & 69.32 & 141.78 & 145.01 & 86.17 \\
    & Noise Ratio : 25\% & \checkmark
    & 114.82 & 94.03 & 58.88 & 97.31 & 72.08 & 45.08 & 87.98 & 88.37 & 68.71 & 140.09 & 144.24 & 85.21 \\
    \cdashline{1-15}[1pt/1pt]

(b) & Image only & \checkmark
    & 131.81 & 109.98 & 74.21 & 113.71 & 87.09 & 59.54 & 105.88 & 103.12 & 85.44 & 152.21 & 154.55 & 92.71 \\
    & Keypoint only & \checkmark
    & 118.21 & 96.12 & 61.71 & 100.87 & 74.87 & 46.91 & 90.12 & 89.21 & 70.77 & 141.64 & 146.21 & 87.88 \\
    \cdashline{1-15}[1pt/1pt]

(c) & HMR2.0~\cite{4dhumans} & \checkmark
    & 149.31 & 125.69 & 80.74 & 100.21 & 89.74 & 56.91 & 104.72 & 104.75 & 94.32 & 134.71 & 176.46 & 132.27 \\
    & BEDLAM-CLIFF~\cite{bedlam, cliff} & \checkmark
    & 133.75 & 100.29 & 73.24 & 103.98 & 83.21 & 54.28 & 92.77 & 96.48 & 75.87 & 147.51 & 166.07 & 126.90 \\
    \cdashline{1-15}[1pt/1pt]

(d) & 160 Tokens & \checkmark
    & 117.38 & 98.12 & 61.94
    & 101.56 & 75.83 & 47.21
    & 90.47 & 90.28 & 71.04
    & 144.78 & 147.91 & 89.63 \\
    & 640 Tokens & \checkmark
    & 127.92 & 107.43 & 64.75
    & 106.67 & 77.69 & 50.36
    & 93.81 & 96.92 & 75.08
    & 149.35 & 151.80 & 94.12 \\
    \cdashline{1-15}[1pt/1pt]

    & \textbf{Ours} & \checkmark
    & \textbf{114.52} & \textbf{93.73} & \textbf{58.01}
    & \textbf{97.02}  & \textbf{71.97}  & \textbf{44.98}
    & \textbf{87.11}  & \textbf{87.91}  & \textbf{68.01}
    & \textbf{139.64} & \textbf{143.74} & \textbf{84.91} \\
    \hline

(e) & Amputation Only (Single) &  
    & 115.70 & 93.75 & 59.08 & 96.32 & 72.76 & 45.92 & 74.71 & 74.51 & 49.93 & 118.09 & 131.12 & 78.08 \\
    & Non Amputation Only (Single) &
    & 113.09 & 92.07 & 57.97 & 95.34 & 72.02 & 45.02 & 76.01 & 76.31 & 50.99 & 120.81 & 134.71 & 81.82 \\
    & \textbf{Ours (Unified)} &
    & \textbf{112.83} & \textbf{91.74} & \textbf{58.62}
    & \textbf{95.26}  & \textbf{71.77}  & \textbf{44.94}
    & \textbf{73.42}  & \textbf{73.19}  & \textbf{49.42}
    & \textbf{116.42} & \textbf{129.25} & \textbf{77.18} \\
    \hline
  \end{tabular}
}
\vspace{-2mm}
\caption{\textbf{Ablation Experiments on the Components of BPAC-Net and AJAHR Tokenizer.} We compare the performance across (a) ablation of BPAC-Net components, (b) evaluation of AJAHR with Gaussian noise-injected keypoints as input BPAC-Net, (c) joint training of baseline methods with BPAC-Net, (d) comparison of AJAHR performance with 160 and 640 tokens and (e) a comparison between single-tokenizer and tokenizer-switching strategies.}
\label{tab:combined_all}
\vspace{-2mm}
\end{table*}

%% file: equtations/total_loss.tex
\begin{equation}
\begin{aligned}
\mathcal{L}_{overall} &= \lambda_{\theta} \mathcal{L}_{\theta} (\boldsymbol{\theta}, \boldsymbol{\hat{\theta}})
+ \lambda_{\beta} \mathcal{L}_{\beta} (\boldsymbol{\beta}, \boldsymbol{\hat{\beta}}) \\
&\quad + \lambda_{2D} \mathcal{L}_{2D} (\mathbf{J}_{2D}, \mathbf{\hat{J}}_{2D})
+ \lambda_{3D} \mathcal{L}_{3D} (\mathbf{J}_{3D}, \mathbf{\hat{J}}_{3D}) \\
&\quad + \lambda_{cls} \mathcal{L}_{cls}.
\end{aligned}
\label{eq:overall_loss}
\end{equation}


%% file: sec/4_experiments_copy.tex
\section{Experiments}
\begin{figure*}[t]
  \centering
  \includegraphics[width=\textwidth]{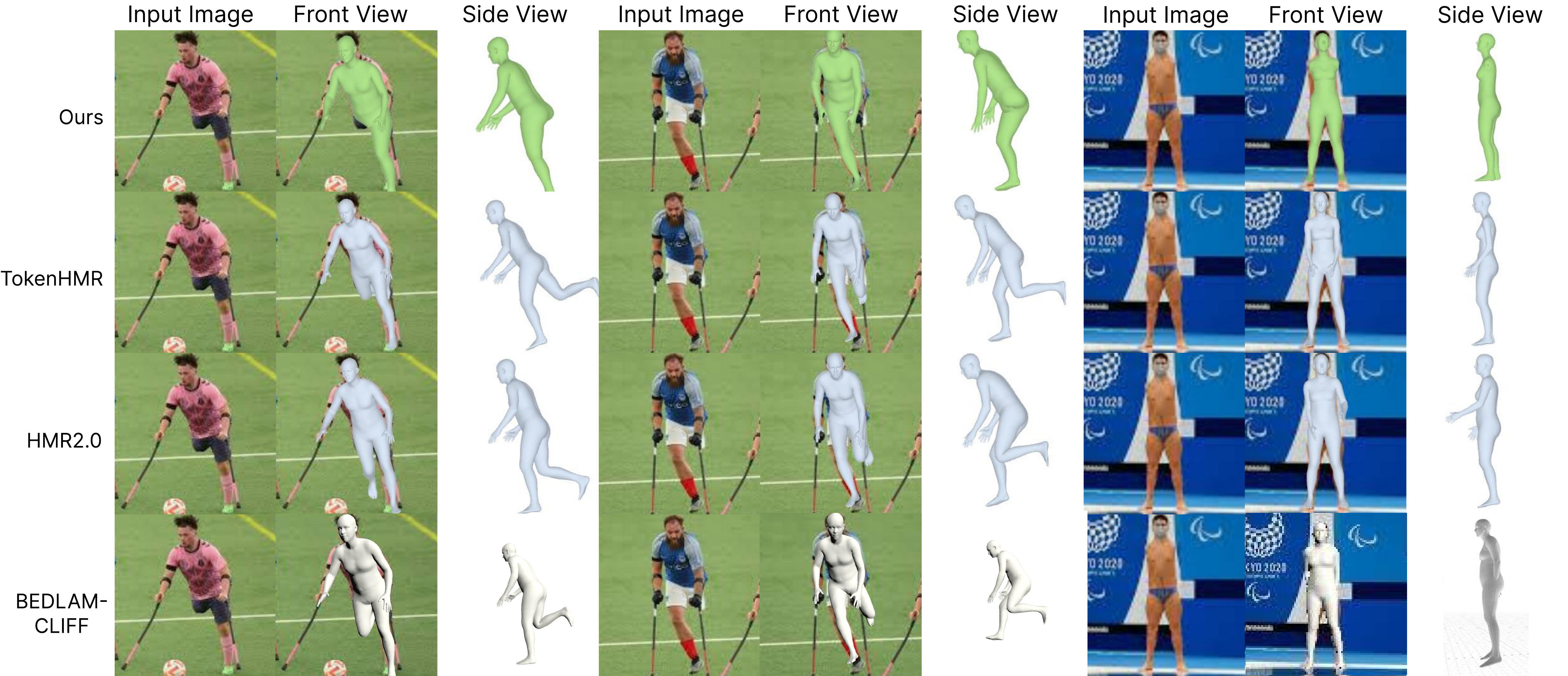}
  \vspace{-5mm}
  \caption{\textbf{A qualitative comparison with other Human Mesh Recovery methods trained on the A3D Dataset.} Unlike TokenHMR~\cite{tokenhmr}, HMR2.0~\cite{4dhumans}, and BEDLAM-CLIFF~\cite{bedlam, cliff}, which do not employ BPAC-Net to identify amputated regions, AJAHR leverages BPAC-Net to explicitly represent these regions in the input images.}
  \label{fig:qualitative}
\end{figure*}

\noindent\textbf{Training Datasets.}
To train the AJAHR-Tokenizer, we used the training split of AMASS~\cite{AMASS}, MOYO~\cite{moyo}, and the pose data from our A3D dataset. Following the TokenHMR~\cite{tokenhmr} training protocol, the classifier and pose‑estimation modules were jointly optimized. For AJAHR training, we employed BEDLAM~\cite{bedlam}, a synthetic dataset with accurate ground-truth 3D annotations, alongside standard datasets used in prior works~\cite{pare, 4dhumans, spin}, including Human3.6M~\cite{h36m_pami}, MPI-INF-3DHP~\cite{3dhp}, COCO~\cite{mscoco}, and MPII~\cite{mpii}. Additionally, similar to HMR2.0~\cite{4dhumans}, we incorporated in-the-wild 2D datasets, such as InstaVariety~\cite{instavariety}, AVA~\cite{ava}, and AI Challenger~\cite{aichallenger}. Furthermore, we included our A3D dataset to enhance training diversity. To ensure a fair comparison in our experiments, we finetuned open-source models, including CLIFF~\cite{cliff}, HMR2.0b~\cite{4dhumans}, and TokenHMR~\cite{tokenhmr}, using the same training data recipe and evaluated their performance against ours. Additional implementation details of AJAHR are provided in the Supplementary Sec. F.

\noindent\textbf{Evaluation Dataset and Metrics.} For tokenizer evaluation, we use Mean Vertex Error (MVE) and Mean Per Joint Position Error (MPJPE). Final pose accuracy is assessed using MVE, MPJPE, and Procrustes-Aligned MPJPE (PA-MPJPE), which calculates the average 3D joint distance (in millimeters) after aligning the estimated and ground-truth joint sets via Procrustes analysis~\cite{procrustes}. Classifier performance is evaluated via confusion matrices, from which we derive metrics such as accuracy, precision, recall, and F1 score. Experiments are conducted on the test splits of 3DPW~\cite{3dpw}, EMDB~\cite{emdb}, and our A3D dataset. To enable fair model comparisons and verify real‑world applicability, we additionally evaluated the models on the In‑the‑Wild Amputee (ITW‑amputee) dataset. This dataset was built by partially synthesizing web‑crawled images using the A3D pipeline and manually annotating them. It includes both single‑ and multi‑limb amputation cases captured in daily-life, rehabilitation, and sports scenes.

\subsection{Robust Generalization to Amputee and Non-Amputee Subjects}
In~\cref{tab:amputation_table}, all evaluation models utilize Ground Truth (GT) labels for amputation regions to remove the corresponding body parts from the estimated mesh before evaluation. The integration of features extracted from BPAC-Net with cross attention results in overall superior performance compared to other human mesh recovery models. Notably, on the ITW-amputee dataset, the proposed method outperforms TokenHMR~\cite{tokenhmr}. This indicates that training on the amputee dataset using the AJAHR-Tokenizer enhances the model’s ability to reconstruct body poses specific to individuals with limb differences. In AJAHR, human mesh recovery is performed in the same manner for both amputee and non-amputee subjects. In~\cref{tab:nonamputation_table} presents the evaluation of human mesh recovery on non-amputee datasets (3DPW~\cite{3dpw}, EMDB~\cite{emdb}). The results demonstrate that the AJAHR model exhibits comparable performance in reconstructing human meshes for non-amputee subjects. This improvement can be attributed to the application of cross attention, which enhances the model’s ability to reconstruct body poses with greater precision, even for non-amputee data. Furthermore, when considered alongside the results in~\cref{tab:amputation_table}, these findings quantitatively confirm that the AJAHR model is effective not only for amputee subjects but also for non-amputee human mesh.

\input{table/3doh50k}
\subsection{Qualitative Experiments}
We conducted a qualitative evaluation on 640 test images from the ITW-amputee dataset, comparing our model against existing models fine-tuned on A3D. While prior methods often failed to reconstruct amputated regions—resulting in distorted overall poses—our model, as shown in~\cref{fig:qualitative}, successfully identified amputation regions even in side-view images and aligned the reconstructed mesh accurately with the person’s location in the image. Notably, in front-view images, existing models frequently misinterpreted amputated legs as folded limbs. In contrast, our model accurately distinguished between amputated and intact limbs, enabling anatomically consistent mesh reconstruction. These results demonstrate its ability to reconstruct amputee-specific body shapes and poses.

\begin{figure*}[h]
  \centering  \includegraphics[width=\textwidth]{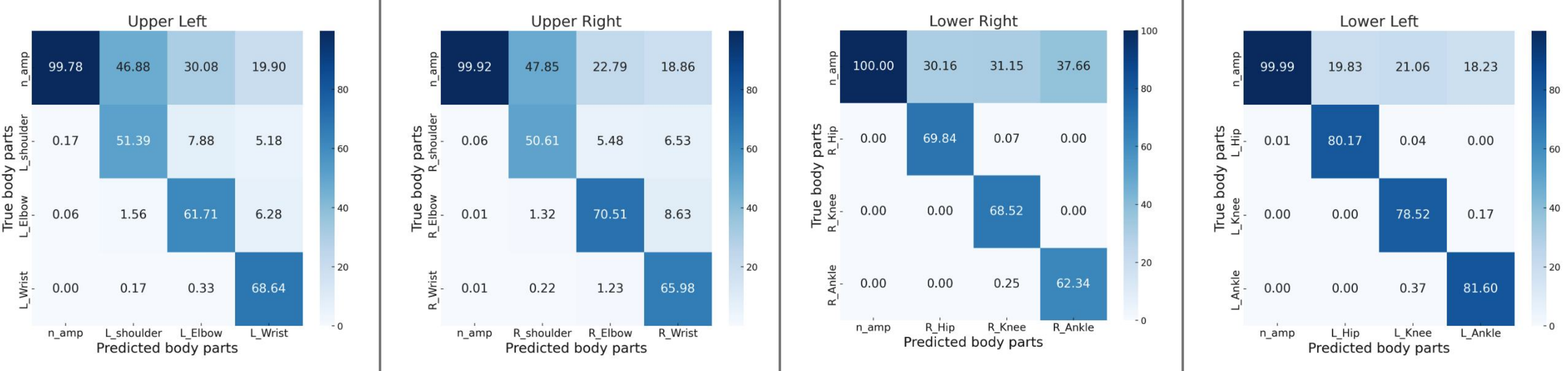}
  \vspace{-6mm}
  \caption{\textbf{Confusion Matrix of A3D Dataset Classification Results using BPAC-Net.} $n_{amp}$ refers to non-amputee, and each label represents the body part predicted by each head. The values are expressed as percentages based on the predicted label columns.}
  \label{fig:confusion}
\end{figure*}

\subsection{Evaluating the Effectiveness of BPAC-Net}
We assess the role of BPAC-Net in enabling amputation-aware mesh recovery through comprehensive experiments. First, we investigate the sensitivity of the overall pipeline to 2D keypoint quality by injecting Gaussian noise at varying levels. As shown in~\cref{tab:combined_all}(a), performance degrades with increasing noise, indicating that 2D detector quality—e.g., ViTPose~\cite{vitpose}—affects both classification and reconstruction. However, the drop remains moderate, suggesting robustness to real-world keypoint imperfections. We then compare single- and multi-modality inputs.~\cref{tab:combined_all}(b) further shows that combining image and keypoint inputs yields superior classification performance over single-modality variants, supporting the effectiveness of our multi-modal design where spatial and appearance cues complement each other. Next, we evaluate BPAC-Net’s generalization to occlusion-heavy scenarios using the 3DOH50K~\cite{3doh50k}. As shown in~\cref{tab:3doh50k}, BPAC-Net distinguishes amputation from occlusion with a high F1 score of 0.977, demonstrating robustness even when limb visibility is affected by occlusion rather than absence. Finally,~\cref{fig:confusion} shows the confusion matrices for each BPAC-Net classifier head. Classification is stable across all body parts, with the $n_{\text{amp}}$ class achieving near-perfect accuracy. The Lower Left and Lower Right heads show fewer misclassifications than their upper-body counterparts, likely due to lower motion variability. These results highlight the effectiveness and reliability of BPAC-Net in conditioning the downstream pose recovery.

\subsection{Necessity of complex model}
As shown in~\cref{tab:combined_all}{(c)}, BPAC-Net built on a weak baseline does not yield performance gains, confirming that reliable pose estimates are essential for effective amputation classification. Furthermore, as presented in~\cref{tab:combined_all}{(e)}, using separate tokenizers for amputees and non-amputees leads to improved performance compared to a unified tokenizer. While full-body tokenization may limit joint-level interpretability, our dual-tokenizer strategy effectively compensates for this limitation by leveraging amputation-aware cues derived from image and keypoint inputs. These results suggest that incorporating structural priors into tokenizer selection is beneficial when modeling subjects with amputation characteristics.

\subsection{Effect of Token Count on AJAHR Performance}
\label{ablation_tokenizer}
\cref{tab:combined_all}(d) presents the results of training the AJAHR model using AJAHR-Tokenizers with 160 and 640 tokens. The results demonstrate that the configuration with 320 tokens (Ours) yields the highest performance. When the number of tokens is insufficient, the model fails to capture pose variations across amputee and non-amputee individuals sufficiently. In contrast, using an large number of tokens introduces redundant information and increases token interference, which leads to degradation in performance.

%% file: table/3doh50k.tex
\begin{table}[t]
\centering
\large
\resizebox{\linewidth}{!}{
  \begin{tabular}{c|cccc|cccc}
    \hline
    \multirow{2}{*}{\centering Method} 
    & \multicolumn{4}{c|}{A3D (amputation)} 
    & \multicolumn{4}{c}{3DOH50K~\cite{3doh50k} (occlusion)} \\
    & Accuracy$\uparrow$ & Precision$\uparrow$ & Recall$\uparrow$ & F1$\uparrow$ 
    & Accuracy$\uparrow$ & Precision$\uparrow$ & Recall$\uparrow$ & F1$\uparrow$ \\
    \hline
    Ours & 0.881 & 0.756 & 0.922 & 0.820
         & 0.956 & 0.956 & 1.000 & 0.977 \\
    \hline
  \end{tabular}
}
\vspace{-2mm}
\caption{\textbf{Amputation Classification Performance on A3D and 3DOH50K.} 
BPAC-Net accurately distinguishes amputation from occlusion, showing consistent classification performance across both synthetic and real-world occlusion scenarios.}
\label{tab:3doh50k}
\vspace{-4mm}
\end{table}

%% file: sec/6_conclusion.tex
\section{Conclusion}
Existing 3D human mesh recovery models~\cite{4dhumans, cliff, tokenhmr} are not designed to handle limb amputations, often hallucinating missing limbs instead of recognizing actual absence. To address this, we present AJAHR—the first framework explicitly built for amputee mesh recovery. Our synthetic data pipeline enables ethical training without rgeal amputee data, and our BPAC-Net-based architecture models amputation explicitly through regional classification and a tokenizer switching mechanism. Extensive experiments show that AJAHR significantly outperforms prior methods on amputee data, while maintaining competitive performance on standard benchmarks.

\noindent\textbf{Limitations and future directions.} AJAHR currently supports only joint-level amputations aligned with the SMPL kinematic tree, and the A3D dataset models only actual amputations, excluding prosthetics. Future extensions will target prosthetic limbs and irregular patterns beyond joint boundaries. From an application standpoint, the framework can support Paralympic sports analysis and inclusive AR/VR systems, enhancing accessibility for individuals with diverse limb differences, such as partial amputations or missing fingers.

%% file: sec/supple.tex
\setcounter{page}{1}
\setcounter{figure}{0}
\setcounter{equation}{0}
\setcounter{table}{0}
\setcounter{section}{0}
\renewcommand{\thefigure}{\Alph{figure}}
\renewcommand{\thetable}{\Alph{table}}
\renewcommand{\theequation}{\arabic{equation}s}
\renewcommand{\thesection}{\Alph{section}}

\twocolumn[{
    \begin{center}
        \vspace{2mm}
        {\Large\bfseries AJAHR: Amputated Joint Aware 3D Human Mesh Recovery\\
        \mdseries Supplementary Material\par}
        \vspace{1.5em}
        {\normalsize
        Hyunjin Cho\textsuperscript{1,3}\textsuperscript{*} \quad
        Giyun Choi\textsuperscript{1}\textsuperscript{*} \quad
        Jongwon Choi\textsuperscript{1,2}\textsuperscript{\dag} \par
        \vspace{0.5em}
        \small
        \textsuperscript{1}Dept. of Advanced Imaging, GSAIM, Chung-Ang University, Korea \quad
        \small
        \textsuperscript{2}Dept. of Artificial Intelligence, Chung-Ang University, Korea \\
        \small
        \textsuperscript{3}Korea Institute of Industrial Technology (KITECH), Korea \\
        {\tt\small \{jincho, cky\}@vilab.cau.ac.kr, choijw@cau.ac.kr}
        \par
        \vspace{0.5em}
        }
        \vspace{2em}
        
    \end{center}
    \begin{center}
    \includegraphics[width=\textwidth]{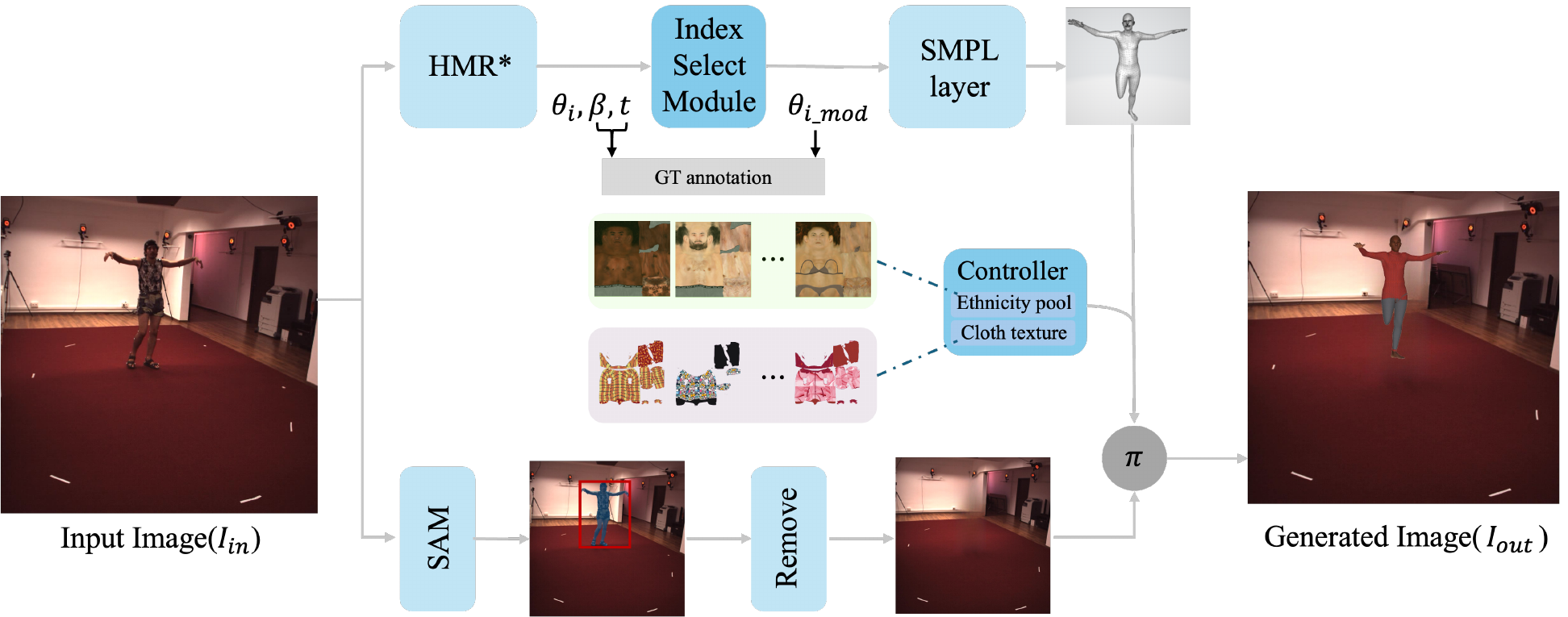}
    \captionof{figure}{\textbf{Pipeline for Synthesizing Images of Individuals with Amputations.} The input image (\(I_{\mathit{in}}\)) is sourced from benchmark datasets~\cite{h36m_pami, mpii, mscoco} and processed through ScoreHMR~\cite{scorehmr}, an HMR-based model (*), which infers SMPL parameters from the image.  We extracted the human region using SAM (Segment Anything Model)~\cite{SAM} for masking and then removed the region using LaMa~\cite{lama} to generate the background. Finally, the generated human mesh was projected onto the background image to create the final image($I_{\mathit{out}}$).}
    \label{fig:pipe_line}
    \end{center}
    \vspace{4mm}
    }]
\blfootnote{* Equal contribution. \quad \textsuperscript{\dag}Corresponding author.}

We provide additional details about AJAHR in the supplementary material.~\cref{sec:pipe_line_supp} introduces the Amputee 3D (A3D) dataset. ~\cref{sec:collapse} explains how we leveraged the SMPL prior to effectively represent amputations on the mesh and details the dataset synthesis pipeline for the A3D Dataset. \cref{sec:zeropose} shows the visual effects of applying non-zero SMPL pose parameters. \cref{sec:a3d_qual} and \cref{sec:proportion} evaluates A3D quality and shows that increasing its proportion improves in-the-wild performance without quality degradation. In \cref{sec:ajahr_id}, \cref{sec:bpac_id}, \cref{sec:tokenizer_id}, introduces detail the architectures, training schemes, and hyperparameters of AJAHR, BPAC-Net, and the AJAHR-Tokenizer, respectively. \cref{sec:ablation_ajahr_moduel} presents quantitative evaluations of the proposed AJAHR model’s tuning strategy and variations in module configuration. 
Lastly, \cref{sec:tokenizer_setting}  presents ablation studies on the number of tokens and codebook size in the AJAHR-Tokenizer.
 
\section{A3D Dataset Synthesis Pipeline Details}
\input{suppl_tex/item}
\label{sec:pipe_line_supp}
The detailed process of the data synthesis pipeline is presented in~\cref{fig:pipe_line}. We employ ScoreHMR~\cite{scorehmr}, which refines predictions by incorporating 2D image cues during inference, making it more effective than conventional regression models for estimating plausible human poses. 

To effectively simulate amputated body parts in the SMPL~\cite{SMPL} representation, we introduce the index select module. This module assigns an index from 0 to 11, representing different amputation types, to the 24 joints of the SMPL body model. The module selects all indices corresponding to the amputated region and its connected child joints, setting their SMPL pose parameters to a zero matrix. This modified SMPL pose is subsequently processed by the SMPL layer, generating a mesh representation of the amputated body. 

In the controller module, we adopt BEDLAM~\cite{bedlam}'s dataset generation approach, which includes incorporating skin and clothing textures to ensure a balanced distribution across two genders (male and female) and seven ethnic groups. This process enhances the diversity of synthetic representations in terms of both ethnicity and appearance. These textures are synthesized onto mannequins in various poses, allowing us to generate images of individuals with diverse limb amputations.

Unlike other works in text-to-image or text-to-motion generation, our approach leverages widely used datasets in the human pose estimation field, such as H36M, MPII, and MSCOCO~\cite{h36m_pami, mpii, mscoco}, ensuring a broad range of diverse poses. Furthermore, by leveraging our synthesis pipeline, we reduce the burden of manually generating diverse poses using motion generation models such as T2M-GPT~\cite{t2m-gpt}, which are commonly used for text-to-motion synthesis. 

The generated amputee meshes are subsequently overlaid onto various background images to improve environmental diversity. To achieve this, we incorporated both indoor lab data and in-the-wild pose images from diverse environments. We further utilized LaMa~\cite{lama} for object removal, where human regions detected by a segmentation model~\cite{SAM} are masked and removed before projecting the amputee mesh onto the cleared area. 

For removing existing humans from the background, we use the Segment Anything Model (SAM)~\cite{SAM} and LaMa~\cite{lama} in sequence: SAM is used to segment the human region, and LaMa inpaints the masked area. Since SAM performs open-vocabulary segmentation based on various prompts (e.g., point clicks, boxes, masks), we prepend a pre-processing module that detects humans via bounding boxes and places point prompts at the top, bottom, left, and right edges of the detected box. This refinement ensures accurate segmentation of only the intended human region and enables clean removal before mesh overlay.

To improve person detection coverage in the input data, we replace Detectron2~\cite{detectron2} with a stronger detector: YOLOv11~\cite{yolo}, fine-tuned on the CrowdHuman dataset~\cite{crowdhuman} for human-only detection. Compared to widely used detectors~\cite{detectron2}, this model detects more individuals across crowded scenes, thereby maximizing the usability of input data and boosting the diversity of generated amputee images.

Finally, we apply a post-processing filter to remove failed or low-quality syntheses. Specifically, we propose a quality checker module that evaluates the realism and visual fidelity of the background after human removal. First, we compute the Structural Similarity Index (SSIM)~\cite{ssim} between the original input image and the LaMa-inpainted background. Images with SSIM scores below 0.5 are excluded from the dataset, as they typically exhibit blurry or overly smoothed artifacts.

Second, we verify whether the human has been successfully removed by applying a 2D human pose detector to the inpainted image. Only images in which no human keypoints are detected are retained. This two-stage filtering process ensures that the background remains both visually natural and free of human remnants, thereby enhancing the overall quality of the synthesized dataset.

\section{Representation of Amputation under the SMPL}
\label{sec:collapse}
In~\cref{fig:smpl_hierarchy}, our study demonstrates that when a specific joint index in the SMPL skeleton hierarchy, such as the R\_Knee joint, is set to a zero matrix and passed through the SMPL Layer, all vertex positions associated with its child joints converge toward the R\_Knee. This behavior effectively illustrates the hierarchical influence of the parent joint on its corresponding child joints within the SMPL model structure.

\begin{figure}[H]
\centering
\includegraphics[width=1\linewidth]{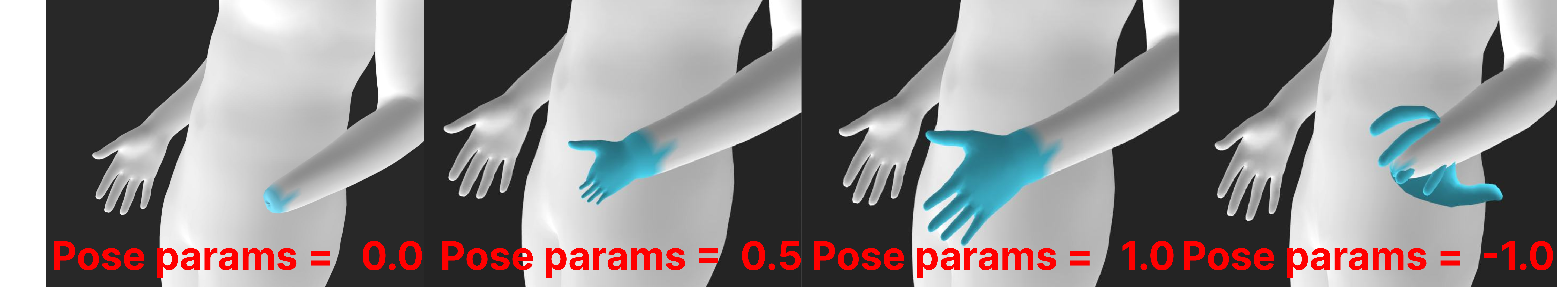}
\caption{\textbf{Effect of varying the pose parameters on the left wrist from $-1.0$ to $1.0$.}}
\label{fig:zeropose}
\end{figure}
\label{sec:zeropose}

\begin{figure*}[h]
  \centering
  \includegraphics[width=\textwidth]{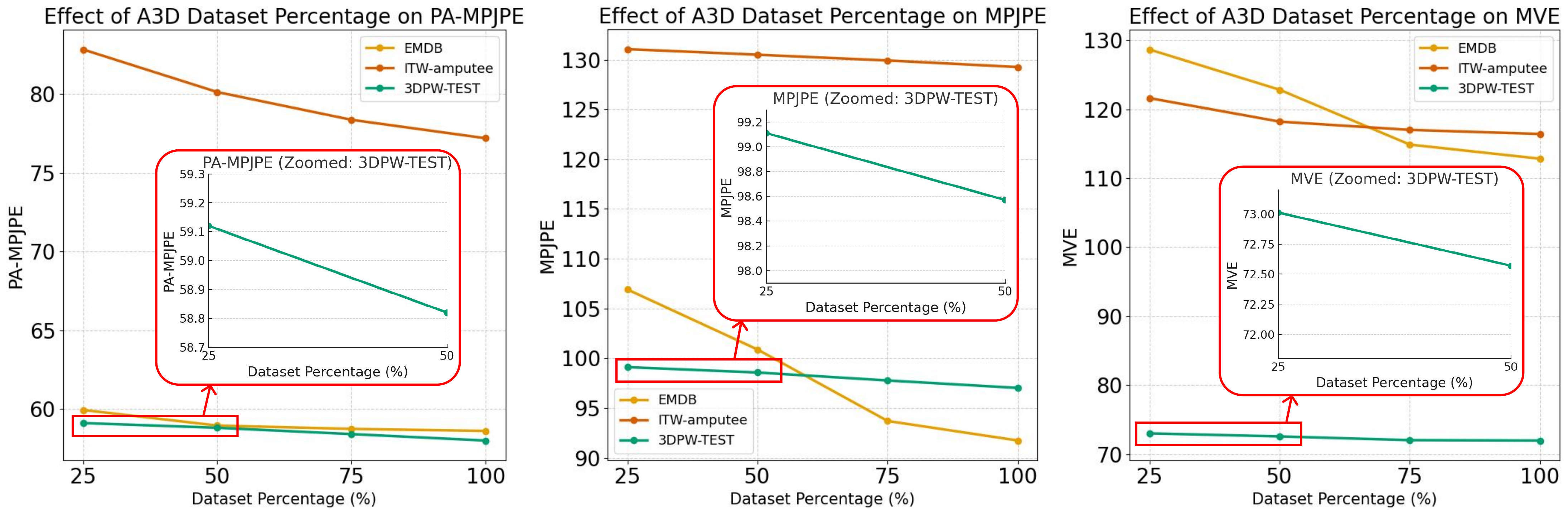}
  \caption{\textbf{Impact of A3D Data Proportion on Performance.} Comparison of model performance when varying the ratio of A3D amputee data within each training batch, evaluated on in-the-wild datasets.}
  \label{fig:dataset_percentage}
\end{figure*}
\input{suppl_tex/sup_bpac}

\section{Visualization with Varying Left Wrist Pose Parameters}
As shown in~\cref{fig:zeropose}, applying a zero pose to the left wrist results in a clean amputation effect, with no distortion in surrounding body parts. In contrast, non-zero values cause visible artifacts or unnatural deformations near the joint. This empirically supports our choice of using zero pose parameters as a reliable proxy for amputated regions while maintaining local shape integrity.

\section{A3D Quality}
\label{sec:a3d_qual}
\vspace{-5mm}
\begin{table}[H]
\small 
\resizebox{\linewidth}{!}{
\begin{tabular}{c|c|c|c|c}
\hline
Dataset     & A3D(MPII~\cite{mpii}) & A3D(MSCOCO~\cite{mscoco}) & A3D(H3.6M~\cite{h36m_pami}) & 
Avg. \\ \hline
LPIPS~\cite{lpips},$\downarrow$ & 0.0735        & 0.0421          & 0.16186        & 0.155         \\ \hline
\end{tabular}
}
\caption{\textbf{LPIPS scores for A3D across datasets.}}
\label{tab:lpips}
\end{table}

\vspace{-3mm}
As shown in \cref{tab:lpips}, our A3D dataset, synthesized as described in \cref{sec:pipe_line_supp}, exhibits high perceptual realism with an average LPIPS~\cite{lpips} of 0.155. Since lower LPIPS corresponds to a smaller perceptual distance between image pairs, this supports that our synthesis pipeline faithfully preserves background, lighting, and texture details at a level comparable to real images. Furthermore, the ablation results in \cref{fig:dataset_percentage} show that simply augmenting the training set with A3D consistently reduces PA-MPJPE and MVE on various in-the-wild benchmarks, including 3DPW~\cite{3dpw} and ITW-amputee. This suggests that A3D sufficiently incorporates outdoor visual factors (such as lighting variation and background complexity) commonly encountered in in-the-wild environments, thereby enabling the model to improve its generalization when trained with it. The high perceptual quality of A3D thus plays a critical role in boosting human mesh recovery performance on natural images even without real amputee data.

\vspace{-3pt}
\input{table/table67_combined}
\section{Effect of A3D Proportion on In-the-Wild Performance}
\label{sec:proportion}
In \cref{fig:dataset_percentage}, the effect of varying the proportion of the amputee dataset A3D within the training batch is examined under in-the-wild evaluation settings. The ratio of A3D was gradually increased from 25\% to 100\% of each training batch, and the model was evaluated at each stage. As the proportion of A3D increased, consistent improvements were observed across all evaluation metrics, including MPJPE, PA-MPJPE, and MVE.

The most notable improvement in PA-MPJPE was observed on the ITW-amputee dataset, whereas EMDB~\cite{emdb} exhibited the most significant gains in MPJPE and MVE. These results suggest that the model effectively learns from the A3D dataset and benefits from increased exposure, resulting in enhanced reconstruction performance for both amputee and non-amputee subjects.

\section{AJAHR Architecture Implementation Detail}
\label{sec:ajahr_id}
AJAHR architecture utilizes ViTPose~\cite{vitpose} as the backbone network to embed input images, and adopts a Transformer decoder~\cite{transformer} following HMR2.0~\cite{hmr}.  
Among the output tokens from the Transformer decoder, the global orientation, body shape, and camera translation are each regressed through separate linear layers.  
For the body pose, in order to match the distribution of AJAHR-Tokenizer, the 1024-dimensional features are passed through six sequential blocks, each composed of two multilayer perceptrons (MLPs) and a GELU~\cite{gelu} activation function.  
We adopt a partially fine-tuned strategy, where only the last four blocks of the ViTPose~\cite{vitpose} backbone, the patch embedding layer, the pose embedding layer, and the final two blocks of the Transformer decoder are updated during training.  

AJAHR is trained in parallel on two NVIDIA A100 GPUs using the AdamW~\cite{adamw} optimizer, with a batch size of 64, a learning rate of $5e^{-6}$, and a weight decay of $1e^{-4}$.  
To ensure balanced learning, we sample the amputee and non-amputee datasets with equal probability (0.5 each).  
Training is conducted for a total of 150{,}000 iterations.  
\vspace{-2mm}
\section{BPAC-Net Architecture Implementation Detail}
\label{sec:bpac_id} The architecture of Body Part Amputation Classifier(BPAC-Net) is presented in~\cref{fig:sup_bpac}, where the input and output feature dimensions of each learnable block are indicated below the respective blocks. BPAC-Net takes batch images along with their corresponding keypoint information as input, which are transformed into 2D keypoint heatmaps before being fed into the ResNet-32 \cite{resnet} backbone. ResNet-32 consists of 16 Basic Blocks, with each block enhanced by a Convolutional Block Attention Module (CBAM)~\cite{woo2018cbam} at its endpoint to perform spatial and channel attention. The final feature vector output from the ResNet-32+CBAM module is 512-dimensional, which is then processed by dedicated classifier heads. Each head predicts one of three amputation types or the non-amputated state, resulting in a 4-dimensional output vector for classification. 
During inference, we replace the Ground Truth (GT) keypoints with 2D keypoints predicted from the image using the ViTPose~\cite{vitpose}.

\section{AJAHR-Tokenizer Implementation Detail}
\label{sec:tokenizer_id}

The AJAHR-Tokenizer architecture is inspired by TokenHMR~\cite{tokenhmr}.  
Both the encoder and decoder consist of four 1D convolutional layers and a single ResNet~\cite{resnet} block.  
It includes a codebook of size $256 \times 2048$ and 320 pose tokens.  
This configuration was selected based on the lowest reconstruction errors in MPJPE and MVE metrics across the AMASS~\cite{AMASS}, MOYO~\cite{moyo}, and A3D datasets. Training is conducted for 200{,}000 iterations with a batch size of 256, a learning rate of $2 \times 10^{-4}$, a gamma value of $5 \times 10^{-2}$, and a weight decay of $1 \times 10^{-5}$.  
To avoid data imbalance, amputee and non-amputee samples are drawn with equal probability (0.5 each) during training.

\section{Ablation Study of AJAHR Training}
\label{sec:ablation_ajahr_moduel}
~\cref{tab:ajahr_moduel_ablation} compares the performance of the AJAHR model across different training strategies. Here, \textbf{Full Fine-Tuning} refers to updating all model parameters except for the frozen AJAHR-Tokenizer, while \textbf{Partially Fine-Tuning} follows the methodology outlined in~\cref{sec:ajahr_id}, where only a subset of parameters is updated. In all experiments, the AJAHR-Tokenizer remains pre-trained and frozen, not updated during training.

In~\cref{tab:ajahr_moduel_ablation}(a), the evaluations utilize Ground Truth (GT) labels of amputation regions, removing the corresponding body parts from the predicted mesh prior to assessment. On the non-amputee dataset, performance differences between Partially Fine-Tuning and Full Fine-Tuning were minimal. However, on the EMDB~\cite{emdb}, Full Fine-Tuning demonstrated superior performance. Notably, the Partially Fine-Tuning approach combined with cross-attention achieved the best performance on the ITW-amputee dataset. This indicates that the partially fine-tuning strategy enabled by cross-attention effectively enhances mesh reconstruction performance irrespective of amputation status.

Meanwhile,~\cref{tab:ajahr_moduel_ablation}(b) extends the experiments by using amputation states predicted by BPAC-Net instead of GT labels. The predicted labels were used to modify SMPL pose parameters, and evaluations were conducted based on meshes reflecting these amputation states.

On amputee datasets (A3D, ITW-amputee), our method recorded the lowest errors in both MPJPE and PA-MPJPE metrics, demonstrating that BPAC-Net reliably identifies amputation sites and significantly improves reconstruction accuracy. Although some cases involved misclassification, incorrectly removing existing body parts or generating non-existent limbs, the overall improvement in performance was clearly evident. These results validate the proposed modular architecture and illustrate the practical capability of BPAC-Net to reliably predict amputation status in real-world scenarios. Furthermore, there was no noticeable performance degradation on non-amputee datasets after applying BPAC-Net, confirming its stability and reliability in general human mesh reconstruction tasks.

\input{table/TOKENIZER}

\section{Ablation Study on AJAHR-Tokenizer Settings}
\label{sec:tokenizer_setting}

As shown in~\cref{tab:tokenizer_qant}, performance improved consistently with increased codebook size, highlighting the importance of adequately large codebooks for representing diverse and complex human structures and poses. However, excessively increasing the number of tokens posed a risk of overfitting without further performance gains. Considering these results, we selected a final tokenizer configuration of a $256 \times 2048$ codebook with 320 tokens for training the AJAHR model, achieving optimal performance.

\paragraph{Acknowledgements.} {\small
This research was partly supported by Culture, Sports and Tourism R\&D Program through the Korea Creative Content Agency grant funded by the Ministry of Culture, Sports and Tourism in 2023 (Project Name : Development of high-freedom large-scale user interaction technology using multiple projection spaces to overcome low-light lighting environments, Project Number : RS-2023-00222280, Contribution Rate : 50\%) and the Institute of Information \& Communications Technology Planning \& Evaluation (IITP) grant funded by the Korea government (MSIT) [IITP-2025-RS-2024-00437102, ITRC(Information Technology Research Center) support program; RS-2021-II211341, Artificial Intelligence Graduate School Program (Chung-Ang University)].}

%% file: suppl_tex/item.tex
\begin{figure}[t]
\centering
\includegraphics[width=0.9\linewidth]{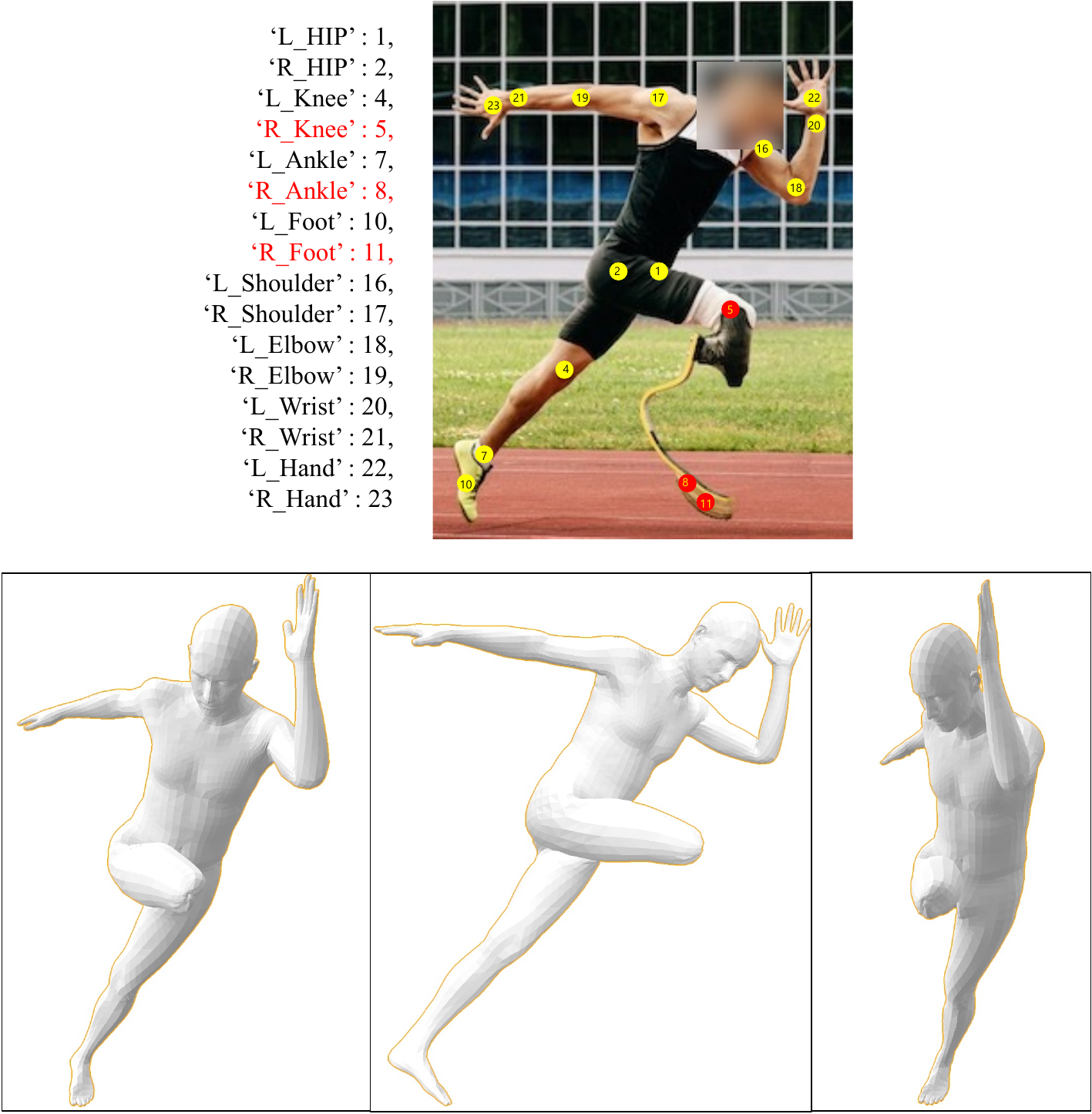}
\vspace{-2mm}
\caption{The top image illustrates the visualization of SMPL~\cite{SMPL} Skeleton Hierarchy and 3D Joint index mapping 
The bottom image presents multiple views of the scenario where the R\_Knee (index 5) is the parent joint, while R\_Ankle (index 8) and R\_Foot (index 11) are its corresponding child joints. Setting the R\_Knee to a zero matrix causes all vertex positions associated with these child joints to converge toward the parent joint, as visualized from different angles.}
\label{fig:smpl_hierarchy}
\end{figure}

%% file: suppl_tex/sup_bpac.tex
\begin{figure*}[t]
\centering
\includegraphics[width=0.95\textwidth]{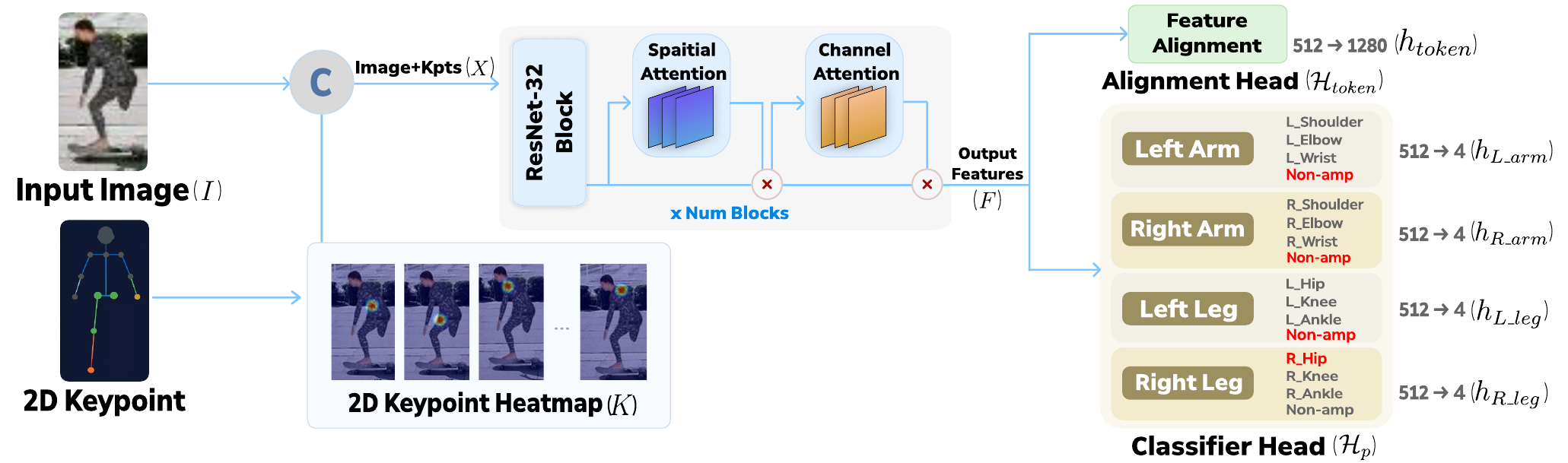}
\vspace{-0.2cm}
\caption{\textbf{Overview of the Body Part Amputation Classifier (BPAC-Net).} The input image \(I\) and 2D keypoints \(K\) (converted to heatmaps) are concatenated and processed through a ResNet-32 \cite{resnet} enhanced with Convolutional Block Attention Module (CBAM) \cite{woo2018cbam}, which applies spatial and channel attention. The extracted features \(F\) are fed into a feature alignment head to produce \(h_{\text{token}}\), which is later used in the Transformer Decoder \cite{transformer} via cross-attention. Four classifier heads \(\mathcal{H}_p \in \{\mathcal{H}_{L_{arm}}, \mathcal{H}_{R_{arm}}, \mathcal{H}_{L_{leg}}, \mathcal{H}_{R_{leg}}\}\), predict amputation status for each corresponding body part.}
\label{fig:sup_bpac}
\end{figure*}

%% file: table/table67_combined.tex
\begin{table*}[t]
\centering
\renewcommand{\arraystretch}{1.1} 
\resizebox{\linewidth}{!}{
\begin{tabular}{c|cccc|ccc|ccc|ccc|ccc}
\hline
\textbf{} & \textbf{Full}         & \textbf{Partially}   & \textbf{Cross}     & \textbf{Use}        & \multicolumn{3}{c|}{\textbf{EMDB}~\cite{emdb}}        & \multicolumn{3}{c|}{\textbf{3DPW}~\cite{3dpw}}         & \multicolumn{3}{c|}{\textbf{A3D}}         & \multicolumn{3}{c}{\textbf{ITW-amputee}}     \\
\textbf{} & \textbf{Fine tuning}  & \textbf{Fine tuning} & \textbf{Attention} & \textbf{Classifier} & MVE$\downarrow$ & MPJPE$\downarrow$ & PA-MPJPE$\downarrow$ & MVE$\downarrow$ & MPJPE$\downarrow$ & PA-MPJPE$\downarrow$ & PVE$\downarrow$ & MPJPE$\downarrow$ & PA-MPJPE$\downarrow$ & PVE$\downarrow$ & MPJPE$\downarrow$ & PA-MPJPE$\downarrow$ \\
\hline
\textbf{(a)} & \checkmark  &             &             &                      & 114.53 & 94.37 & 58.68 & 97.16 & 72.68 & 46.00 & 74.98 & 74.45 & \textbf{48.80} & 121.62 & 133.73 & 77.94 \\
~            &             & \checkmark  &             &                      & 112.99 & 92.05 & 58.84 & 95.73 & \textbf{71.33} & 47.72 & 73.98 & 74.30 & 49.71 & 122.34 & 132.95 & 81.23 \\
~            & \checkmark  &             & \checkmark  &                      & 116.78 & 95.84 & \textbf{57.29} & 98.66 & 72.01 & 44.95 & 75.99 & 73.29 & 49.54 & 126.46 & 139.05 & 83.88 \\
~            &             & \checkmark  & \checkmark  &                      & \textbf{112.83} & \textbf{91.74} & 58.62 & \textbf{95.26} & 71.77 & \textbf{44.94} & \textbf{73.42} & \textbf{73.19} & 49.42 & \textbf{116.42} & \textbf{129.25} & \textbf{77.18} \\
\cdashline{1-17}[1pt/2pt]
\textbf{(b)} & \checkmark            &                      &                    & \checkmark          & \textbf{113.42} & 97.33             & 59.87                & 97.98            & 73.68             & 47.76                & 89.34           & 89.96             & 69.87                & 143.49          & 154.83            & 85.17 \\
~            &                      & \checkmark           &                    & \checkmark          & 113.93          & 94.12             & 58.87                & 98.51            & 73.88             & 48.90                & 89.12           & 88.12             & 68.17                & 145.01          & 153.17            & 90.12 \\
~            & \checkmark            &                      & \checkmark         & \checkmark          & 115.65          & 97.98             & 59.08                & 99.34            & 72.23             & 45.16                & 88.51           & 88.32             & 68.19                & 148.28          & 157.51            & 93.11 \\
~            &                      & \checkmark           & \checkmark         & \checkmark          & 114.52          & \textbf{93.73}    & \textbf{58.01}       & \textbf{97.02}   & \textbf{71.97}    & \textbf{44.98}       & \textbf{87.11}  & \textbf{87.91}    & \textbf{68.01}       & \textbf{139.64} & \textbf{143.74}   & \textbf{84.91} \\
\hline
\end{tabular}
}
\caption{\textbf{Comparison of AJAHR Module Ablation Studies.} (a) Partially Fine-Tuning method freezes the
AJAHR parameters while selectively updating a limited set of trainable parameters within each module. In contrast, Full Fine-Tuning
updates all parameters of AJAHR during training. (b) After BPAC-Net infers the body part status, the
corresponding label is used to force the inferred SMPL body pose parameters to zero. The evaluation is then conducted using 3D keypoints
obtained from the reconstructed mesh.
}
\label{tab:ajahr_moduel_ablation}
\end{table*}

%% file: table/TOKENIZER.tex
\begin{table}[t]
\centering
\large
\resizebox{\columnwidth}{!}{
\begin{tabular}{cc|cc|cc|cc}
\hline
\multicolumn{2}{c|}{\textbf{Method}} 
  & \multicolumn{2}{c|}{\textbf{A3D}} 
  & \multicolumn{2}{c|}{\textbf{AMASS~\cite{AMASS}}} 
  & \multicolumn{2}{c}{\textbf{MOYO~\cite{moyo}}} \\ 
\multicolumn{2}{c|}{} 
  & MPJPE$\downarrow$ & MVE$\downarrow$ 
  & MPJPE$\downarrow$ & MVE$\downarrow$ 
  & MPJPE$\downarrow$ & MVE$\downarrow$ \\ 
\hline
CodeBook & 128x2048       & 1.72 & 8.05 & 2.24 & 8.62 & 7.49 & 16.50 \\  (320 tokens) 
         & 256x1024 & 1.92 & 8.50 & 2.50 & 8.80 & 7.15 & 16.18 \\
\hline
\multirow{5}{*}{\shortstack{Tokens\\(Codebook: 256x2048)}} 
  & 20   & 5.51 & 12.60 & 7.62 & 15.00 & 23.10 & 38.87 \\
  & 40   & 3.49 & 9.76  & 4.51 & 10.67 & 14.50 & 26.04 \\
  & 80   & 2.39 & 8.56  & 3.07 & 8.97  & 10.05 & 19.44 \\
  & 160  & 2.07 & 8.38  & 2.74 & 8.67  & 6.68  & 14.60 \\
  & 640  & 2.59 & 9.03  & \textbf{2.60} & 9.03 & 8.02 & 16.60 \\
\hline
\multicolumn{2}{c|}{\textbf{Ours (256x2048, 320 tokens)}} 
  & \textbf{1.56} & \textbf{8.01} & \textbf{1.90} & \textbf{8.08} & \textbf{5.52} & \textbf{13.47} \\
\hline
\end{tabular}
}
\vspace{-1mm}
\caption{\textbf{Ablation Studies on AJAHR Tokenizer Configuration.} The MOYO~\cite{moyo} validation dataset was used for evaluation. In the codebook size comparison experiment, the total number of tokens was set to 320. In the token count comparison, the codebook size was fixed to $256 \times 2048$.}
\label{tab:tokenizer_qant}
\end{table}